\documentclass[Afour,sageh,times]{sagej}

\usepackage{moreverb,url}

\usepackage[colorlinks,bookmarksopen,bookmarksnumbered,citecolor=red,urlcolor=red]{hyperref}
\usepackage{arydshln}
\usepackage{algorithm}
\usepackage{algorithmic}
\usepackage{amsmath}
\usepackage{xifthen}
\usepackage{graphicx}
\usepackage{amsmath}
\usepackage{amsfonts}
\usepackage{amssymb}
\usepackage{amstext}
\usepackage{amsbsy}
\usepackage{epsfig}
\usepackage{epsf}
\usepackage{multirow}
\usepackage{float}
\usepackage{stmaryrd} 

\usepackage{xspace}
\usepackage{url}
\usepackage{verbatim}
\usepackage{ifthen}
\usepackage{rotating}
\usepackage{mathpazo}
\usepackage{bbding}
\usepackage{booktabs}

\newcommand{\sphere}{\ensuremath{\mathcal{S}}\xspace}
\newcommand*{\domain}{\ensuremath{\mathcal{D}}\xspace}
\newcommand{\unit}[1]{\,\mathrm{#1}}
\newcommand{\afluxunit}{\,\ensuremath{\unit{cm}^{-2}\unit{s}^{-1}\unit{st}^{-1}}\xspace}
\newcommand{\direction}{\ensuremath{\vec{\Omega}}\xspace}
\newcommand{\grad}{\vec{\nabla}}

\newcommand{\Sigt}[1][]{\ensuremath{\Sigma_{\xslabel[#1]{t}}}\xspace}
\newcommand{\xslabel}[2][]{\ifthenelse{\isempty{#1}}{\mathrm{#2}}{\mathrm{#2},#1}}
\newcommand{\Sigs}[1][]{\ensuremath{\Sigma_{\xslabel[#1]{s}}}\xspace}
\newcommand{\position}{\ensuremath{\vec{x}}\xspace}
\newcommand{\domg}{\dx[\Omega]}
\newcommand{\dx}[1][x]{\,d#1}
\newcommand{\Sigf}[1][]{\ensuremath{\Sigma_{\xslabel[#1]{f}}}\xspace}
\newcommand{\bnormal}{\ensuremath{\normal_\mathrm{b}}\xspace}
\newcommand{\normal}{\ensuremath{\vec{n}}\xspace}
\newcommand{\omgdnb}{\direction\cdot\bnormal}
\newcommand{\sfluxunit}{\,\ensuremath{\unit{cm}^{-2}\unit{s}^{-1}}\xspace}

\newcommand\BibTeX{{\rmfamily B\kern-.05em \textsc{i\kern-.025em b}\kern-.08em
T\kern-.1667em\lower.7ex\hbox{E}\kern-.125emX}}

\begin{document}

\runninghead{Fande Kong}

\title{Parallel memory-efficient all-at-once algorithms for the sparse matrix triple products in multigrid methods}

\author{Fande Kong\affilnum{1}}

\affiliation{\affilnum{1} Computational Framework, Idaho National Laboratory}

\corrauth{Fande Kong, 2525 Fremont Ave, Idaho Falls, ID 83402}

\email{fande.kong@inl.gov}

\begin{abstract}
Multilevel/multigrid methods  is one of the most popular approaches  for solving a large sparse  linear  system of equations, typically, arising from the discretization of partial differential equations.  One critical step in the multilevel/multigrid methods  is to form  coarse matrices  through a sequence of sparse  matrix triple products. A commonly used approach for the triple products  explicitly  involves two steps, and during each step a sparse matrix-matrix multiplication is employed. This approach works well for many applications with a good computational efficiency, but it has a high memory overhead since some auxiliary  matrices  need to be temporarily  stored for accomplishing  the calculations.  In this work, we propose two  new algorithms that construct  a coarse matrix with taking one pass through the input matrices without involving any auxiliary matrices for saving memory. The new approaches  are referred to as ``all-at-once" and ``merged all-at-once" (a modified version of ``all-at-once") since the new algorithms  calculate  the two sparse matrix-matrix multiplications simultaneously, and the traditional method  is denoted as ``two-step".  The all-at-once and the merged all-at-once algorithms are implemented based on hash tables in PETSc as part of this work with a careful consideration on the performance in terms of the compute time and the memory usage.  In the new methods, the first sparse matrix-matrix multiplication is implemented using a row-wise algorithm, and the second one is based on an outer  product.  We numerically show that the proposed algorithms and their implementations are perfectly  scalable in both the compute time and the memory usage  with up to $32,768$ processor cores for a model problem  with 27 billions of unknowns. The scalability  is also demonstrated  for a realistic neutron transport problem with over 2 billion unknowns   on a supercomputer with $10,000$ processor cores. Compared with the traditional two-step method, the  all-at-once and the merged all-at-once algorithms consume much less memory  for both the model problem and the realistic neutron transport problem meanwhile they  are able to maintain     the computational efficiency.
\end{abstract}


\maketitle

\section{Introduction}
Multigrid/multilevel methods  including geometric multigrid (GMG) and algebraic multigrid (AMG) is one of the most popular approaches  for solving a large sparse  linear system of equations, $Ax=b$, arising from the discretization  
of partial different equations (\cite{smith2004domain, stuben2000algebraic}). The methods involves  generating  a sequence of linear systems of decreasing size during the setup phase, and iteratively improving the solution to the original system using the sequence of coarse systems in the solve phase.  One critical component of the multigrid/multilevel methods is to form a coarse operator $C$ through a sparse matrix triple product of  restriction $P^T$,  fine operator $A$,  and interpolation $P$.  Here  $P^T$ is  the transpose of $P$ when using the Galerkin method, and $P$ can be generated  either geometrically (\cite{kong2016highly, kong2017scalable, kong2018scalability, kong2018simulation}) or 
algebraically (\cite{kong2018fully, kong2019highly}).  It is challenging to design  an efficient parallel  algorithm and develop  its corresponding software  for the sparse matrix triple product since
we have to consider both the memory efficiency and the compute time efficiency.   Beside the multigrid/multilevel methods,  the sparse matrix triple product is also used in other areas  such as the Schur complement algorithms so that developing an efficient triple product algorithm  becomes an active research topic.

Let us briefly  review a few of these developments, and interested readers are referred to (\cite{ballard2016reducing, bulucc2012parallel, mccourt2013efficient}) for more literature reviews. Many previous works have considered parallel algorithms for sparse matrix-matrix multiplications  for the general-purpose use (\cite{akbudak2014simultaneous}) and the particular application (\cite{challacombe2000general}). In (\cite{ballard2016reducing}), the authors propose and analyze a sparse SUMMA algorithm that was originally used for dense matrices. (\cite{borvstnik2014sparse}) uses a similar idea to convert a dense matrix  algorithm (Cannon) to its parallel sparse version for quantum-chemical applications with special optimizations and tuning.  
In the context of the multigrid/multilevel methods, (\cite{tuminaro2000parallel}) concerns on the smoothed aggregation algebraic multigrid on distributed-memory supercomputers, where a row-wise algorithm is used for the sparse matrix-matrix multiplications of the two-step  matrix triple product.  (\cite{mccourt2013efficient}) proposes a matrix coloring technique to cast a sparse matrix-matrix multiplication to a sparse-matrix dense-matrix operation, and the method is used in the multigrid/multilevel methods.   (\cite{ballard2016hypergraph}) studies a hyper graph  to represent the communication costs of the parallel  algorithms for general sparse matrix-matrix multiplications, where a Gakerkin  triple product is employed as a case study, and the authors conclude that the row-wise algorithm is communication efficient for the first  matrix-matrix multiplication, but inefficient for the second one. 

In most of the literatures, typically, the sparse matrix triple product, $C=P^TAP$, is formed using two separate  sparse matrix-matrix multiplications, $A \cdot P$ and $P^T \cdot (AP)$ (\cite{ballard2016hypergraph, ballard2016reducing, petsc-user-ref}). This two-step method works well for many applications, but it is difficult to use the two-step approach  for  certain  memory-intensive     applications such as neutron transport problems (\cite{kong2018fully, kong2019highly}) since the two-step method needs to create  some auxiliary matrices  causing   a high memory overhead in order to efficiently cary out the calculations. To overcome this difficulty, we develop and study two all-at-once algorithms (including a merged version) that form the coarse operator $C$ with one pass through $P^T$, $A$ and $P$ without creating any auxiliary matrices. Here a row-wise algorithm is employed for the first matrix-matrix multiplication, and an outer  product is adopted for the second matrix-matrix multiplication. Note that in the new all-at-once approaches,  the two matrix-matrix multiplication are carried out  simultaneously, which will be  discussed for more details in Section 3.    Compared with the traditional two-step method, the new all-at-once algorithms  and their implementations  result in a reduction of the memory usage by a factor of $9\times$ for a model problem and a factor of $2.5\times$ for a realistic neutron transport application, meanwhile the new  all-at-once approaches  are able to maintain a good computational  efficiency.  We numerically show that the proposed algorithms  work efficiently for both the model problem and the realistic problem with up to 27 billions of  unknowns  on supercomputers using up to  $32,768$ processor cores.

The rest of this paper is organized as follows. In Section 2, a row-wise matrix-matrix multiplication  that serves 
as part of the calculations in a triple product   is described
in detail. Two new all-at-once algorithms based on hash tables for the sparse matrix triple products in multigrid/multilevel methods  are described in
Section 3, and numerical results for a model problem and a realistic neutron transport problem with up to $32,768$ processor cores  are
given and discussed  in Section 4. The results are summarized and conclusions are drawn in Section 5.

\section{Sparse matrix-matrix multiplications}
There are different types of  algorithms for  sparse matrix-matrix multiplications. The algorithms are generally classified into 1D, 2D  and 3D  according to the partitioning schemes used to assign the computations  of matrix  to each processor core. Based on the current infrastructures in PETSc, we consider 1D algorithms only in this work. In the 1D approaches, each processor core owns a chunk of the rows of matrix using  a row partition or  a chunk of  the columns of matrix using  a column partition. The 1D matrix-matrix multiplication algorithms are further divided into row-wise, column-wise, and  outer   product. A matrix in PETSc is distributed  in  row wise so that a row-wise algorithm is  chosen. Next we describe the row-wise algorithm that is employed in the first matrix-matrix multiplication of our new all-at-once approaches.  

Let $A \in \mathbb{R}^{n,n}$, and $P \in \mathbb{R}^{n,m}$, where $n$ is  associated  to the number of unknowns on the fine mesh, and $m$ is determined by the coarse mesh.  A matrix-matrix multiplication operation is defined such that the $j$th column of the $i$th row of output matrix $C$, denoted as $C(i,j)$, is calculated as follows
$$
C(i,j) =  \sum_{k=1}^{n} A(i, k) P(k, j).
$$
The atomic task in the row-wise algorithm is the calculation  of a row of $C$, that is, 
$$
C(i,:) =  \sum_{k=1}^{n} A(i, k) P(k, :).
$$
Here the $i$th row of $C$, $C(i,:)$, is a linear combination of  the rows of $P$, some of which are not local, corresponding to the nonzero columns of  the $i$th row of $A$.  In the row-wise algorithm, 
the matrix is equally  distributed to processors in block rows, where each processor owns $n_l = n/(np)$ consecutive rows shown in Eq.~\eqref{eq:partition}.   Here $np$ is the number of processor cores. For  simplicity of discussion, we assume $n$ is divisible by $np$. 
In reality, $n$ may be not divisible by $np$, and if it is the case, some processor cores will own a few more rows than others.  Let $A_l$ denote the rows of $A$ owned by the $l$th processor core, that is, 
$$
A_l = A(ln_l: (l+1)n_l-1 ,:).
$$ 
For  simplicity  of description, we also define the $p$th block columns of $A_l$ as $A_{lp} = A_l(:, pn_l: (p+1)n_l-1)$ as shown in Eq.~\eqref{eq:partition}. 
\begin{equation}\label{eq:partition}
A =
\left [
\begin{array} {llll}
A_1 \\
\hline
A_2\\
\hline
\cdots\\
\hline
A_{np}\\
\end{array}
\right ]
=
\left [
\begin{array} {c:c:c:c}
A_{11} & A_{12} & \cdots & A_{1,np} \\
\hline
A_{21} & A_{22} & \cdots & A_{2,np} \\
\hline
\vdots & \vdots & \cdots &  \vdots \\
\hline
A_{np,1} & A_{np,2} & \cdots & A_{np,np} 
\end{array}
\right ]
\end{equation}
 Note that $A_{lp}$ is owned by the $l$th processor core, and represents the relationship with the $p$th 
processor core. If $A_{lp} = 0$, there is no communication from processor $p$ to $l$.  A good partition of $A$ should make most of submatrices $A_{lp}$ zero  to minimize the communication.  We do not partition $A$ directly, instead,   the partition of $A$ is derived from the finite element mesh partition that is implemented using the approaches in  (\cite{karypis1997parmetis, kong2018general}) since $A$ arises  from the discretization of partial differential equations on a finite  element mesh.  $P$ is also  partitioned in the same way, that is, $P_l = P(ln_l: (l+1)n_l-1, :)$, and $P_{lp} = P_l(:, pm_l: (p+1)m_l-1)$, where $m_l= m/(np)$.  For the $l$th processor core, the computation of the local matrix $C_l$ (the output matrix $C$ has the same row partition as $A$, and the same column partition  as $P$) is written as 
\begin{equation}\label{eq:localmatrixC}
C_l =  \sum_{p=1}^{np}  A_{lp} P_p.
\end{equation}
As stated earlier, some of submatrices do not have nonzero elements  since $A$ is a sparse matrix,  and therefore the calculation of  $C_l$ does  not involve all processor cores, instead, only communicates  with the processor cores corresponding to nonzero submatrices. Define the number of the nonzero submatrices of $A$ as $n_\text{nzm}$, and  the index function as $\text{I}(k)$ that returns the index of  the $k$th nonzero submatrix of $A$. Eq.~\eqref{eq:localmatrixC} is reformulated as
\begin{equation}\label{eq:localmatrixC_1}
C_l =  \sum_{k=1}^{n_\text{nzm}}  A_{l{\text{I}(k)}} P_{\text{I}(k)}.
\end{equation}
To explain this clearly,  we  present a simple example shown in Eq.~\eqref{eq:rowwise}. 
\begin{equation}\label{eq:rowwise}
\begin{array}{lll}
\left [
\begin{array} {c:c:cc}
 \color{red} * &   \color{red} * & ~ &   \color{red}  * \\ 
~ & * & * & *   \\ 
\hline
* & * & * & *  \\
~ & * & * & ~   \\
\hline
~ & * & * & *  \\
~ & * & *  & *   \\
\end{array}
\right ]
= \\
~\\
\left [
\begin{array} {cc:cc:cc}
 \color{green}  * &   \color{green}  * & ~ & ~ &  \color{green}  * & ~ \\ 
~ & * & *& ~& * & ~   \\ 
\hline
* & ~ & ~ & * & * & ~  \\
~ & * & ~ & * & ~ & ~\\
\hline
~ & ~ & ~ & *  & * & ~  \\
~ & * & ~  & ~ & * & *  \\
\end{array}
\right ]
\left [
\begin{array} {c:c:cc}
\color{blue} * & ~ & ~ & \color{blue} * \\ 
~ &  \color{blue} * & ~  & ~   \\ 
\hline
~ & ~ & * & *  \\
~ & ~ & * & ~ \\
\hline
~ & \color{blue}* & ~ & \color{blue} *  \\
~ & ~ & *  & ~  \\
\end{array}
\right ].
\end{array}
\end{equation}
Here the horizontal  lines represent  the row partition of matrix, and the vertical dashed line is the virtual column partition that matches the row partition of the right matrix if it exists.  For instance,  in the matrix $A$ (the first matrix on the right side of Eq.~\eqref{eq:rowwise}), the first processor core takes the first and the second rows, the second processor core owns the third and the forth rows, and the third processor core has the fifth and the sixth rows. 
In this toy example,  the first row of $C$  is the linear combination of the first, the second, and the fifth rows of $P$ since the first, the second and the fifth columns of $A$ are nonzero.  To calculate the first row of $C$, a remote row of $P$, the fifth row, is needed,  which is implemented by communication.  The calculation of different rows of local $C_l$,  different remote rows of $P$ are involved.  If we did a message exchange 
for each   row calculation individually, it would be inefficient. Thus, we extract all the required  remote rows (forming a matrix  $\tilde{P}_{\text{r}}$) that  corresponds to nonzero columns of $A_{lp}$ ($l \neq p$)   up front.  For example, in Eq.~\eqref{eq:rowwise}, the third, the fifth remote rows of $P$ are extracted for the first processor core because the nonzero off-processor columns of $A_l$ are  the third   and the  fifth columns. In PETSc (\cite{petsc-user-ref}), the local part of matrix, e.g., $A_l$, $C_l$ and $P_l$, is physically implemented as  two blocks, diagonal block (e.g., $A_{\text{d}} = A_{ll}$) and off-diagonal block (e.g., $A_{\text{o}} =\cup_{k=1}^{ k \neq l} A_{lk}$), which  are stored in two sequential matrices in a compressed sparse row (CSR) format.  We here drop the subindex  $l$ for $A_{\text{d}}$ and $A_{\text{o}} $ without confusion.  The formulation of the calculations of  $C_l$ is  rewritten as follows
$$
C_l= A_{\text{d}} P_l + A_{\text{o}} \tilde{P}_r. 
$$ 
The calculation of the matrix-matrix multiplication is generally split into two parts, symbolic calculation and numeric calculation.  In the symbolic calculation, the task is to accurately  preallocate the 
memory for the output matrix $C_l$ with going  through the matrices $A_l$, $P_l$ and $\tilde{P}_{\text{r}}$ without involving  any floating-point operations. In the numeric process, the actual numeric calculation  is carried out and the results are added to the  allocated  space   of $C_l$.  The symbolic calculation is summarized in Alg.~\ref{alg:onerowofAP} and~\ref{alg:symbolicAP}. 
\begin{algorithm}
\caption{Symbolic calculation of one row of AP.\label{alg:onerowofAP}}
\begin{algorithmic}[1]
\STATE{Input: $i$, $A= [A_\text{d}, A_\text{o} ]$, $P_l = [P_{\text{d}}, P_{\text{o}}]$, $\tilde{P}_r$}
\STATE{Initialize $\{R_\text{d}, R_\text{o}\} = \{\emptyset, \emptyset \}$}
\FOR { each nonzero column $k$ in $A_{\text{d}}(i,:)$}
\FOR { each nonzero column $j$ in $P_{\text{d}}(k, :)$}
\STATE Insert $j$ into $R_{\text{d}}$
\ENDFOR
\FOR { each nonzero column $j$ in $P_{\text{o}}(k, :)$}
\STATE Insert $j$ into $R_{\text{o}}$
\ENDFOR
\ENDFOR
\FOR { each nonzero column $k$ in $A_\text{o}(i,:)$}
\FOR { each nonzero column $j$ in $\tilde{P}_{\text{r}}(k, :)$}
\IF {$j$ is a diagonal column}
\STATE Insert $j$ into $R_\text{d}$
\ELSE
\STATE Insert $j$ into $R_\text{o}$
\ENDIF
\ENDFOR
\ENDFOR
\STATE{Output: $\{R_\text{d}, R_\text{o}\}$}
\end{algorithmic}
\end{algorithm}
In Alg.~\ref{alg:onerowofAP}, $R_\text{d}$ and $R_{\text{o}}$ mathematically are integer sets, and they can be implemented based on different data  structures such as linked  list, binary tree, hash table, etc. In this work, we use the hash table to implement  $R_\text{d}$ and $R_{\text{o}}$  for our new all-at-once algorithms  since the hash table has an average $O(1)$ lookup time and also simplifies the implementation. Other implementations  such as the linked  list is also available in PETSc. For simplifying the description, we  assume that  the hash table is used during the following discussion.
\begin{algorithm}
\caption{Row-wise algorithm for symbolic calculation of $AP$. \label{alg:symbolicAP}}
\begin{algorithmic}[1]
\STATE{Input: $A_l=[A_\text{d}, A_\text{o} ]$, $P_l$
\STATE Extract the remote rows, $\tilde{P}_{\text{r}}$, of $P$  corresponding to the nonzero columns of $A_{\text{o}}$}
\STATE Initialize $\text{nzd} = \{0\}$ and $\text{nzo}=\{0\}$
\STATE{Initialize $\{R_\text{d}, R_\text{o}\} = \{\emptyset, \emptyset \}$}
\STATE{$i=1$}
\FOR {$i<=$ the number rows of $C_l$}
\STATE $\{R_\text{d}, R_\text{o}\}$ = Alg.~\ref{alg:onerowofAP}($i$, $A_l$, $P_l$, $\tilde{P}_{\text{r}}$)
\STATE $\text{nzd}(i) = |R_\text{d}|$
\STATE $\text{nzo}(i) = |R_\text{o}|$
\STATE Clear $\{R_\text{d}, R_\text{o}\}$
\STATE $i += 1$
\ENDFOR 
\STATE {Preallocate memory for $C_l$ using $\text{nzd}$ and $\text{nzo}$}
\STATE{Output: $C_l$ with allocated space}
\end{algorithmic}
\end{algorithm}
In Alg.~\ref{alg:symbolicAP},  $\text{nzd} = \{0\}$ and $\text{nzo}=\{0\}$ are integer arrays to store the numbers of nonzero columns for the diagonal and the off-diagonal parts of $C_l$, respectively.  $|\cdot|$ represents the number of elements in a set. The memory of $R_d$ and $R_o$ could be  reused for each row of $AP$, and ``clear" simply resets  a flag in the data structure so that the memory is ready for next row.  Note that line 2 of Alg.~\ref{alg:symbolicAP} involves one message exchange, and all other parts are proceeded  independently for each processor core.  The numeric calculation of the matrix-matrix multiplication using the row-wise algorithm is shown in Alg.~\ref{alg:NumericOneRowOfAP} and~\ref{alg:NumericAP}. 
\begin{algorithm}
\caption{Numeric calculation of one row of AP.\label{alg:NumericOneRowOfAP}}
\begin{algorithmic}[1]
\STATE{Input: $i$, $A_l= [A_{\text{d}}, A_{\text{o}}]$, $P_l=[P_\text{d}, P_{\text{o}}]$, $\tilde{P}_{\text{r}}$}
\STATE{Initialize $R = \emptyset $}
\FOR { each nonzero column $k$ in $A_{\text{d}}(i,:)$}
\FOR { each nonzero column $j$ in $P_{\text{d}}(k, :)$}
\STATE $R(j) +=  A_{\text{d}}(i,k)*P_{\text{d}}(k,j)$
\ENDFOR
\FOR { each nonzero column $j$ in $P_{\text{o}}(k, :)$}
\STATE $R(j) +=  A_{\text{d}}(i,k)*P_{\text{o}}(k,j)$
\ENDFOR
\ENDFOR
\FOR { each nonzero column $k$ in $A_{\text{o}}(i,:)$}
\FOR { each nonzero column $j$ in $\tilde{P}_{\text{r}}(k, :)$}
\STATE $R(j) += A_{\text{o}}(i,k)*\tilde{P}_{\text{r}}(k,j)$
\ENDFOR
\ENDFOR
\STATE{Output: $R$}
\end{algorithmic}
\end{algorithm}
In Alg.~\ref{alg:NumericOneRowOfAP},  $R$ is a hash table that  associates  a key $k$ with its numeric value. ``$+=$" represents that if $j$ already exists in $R$ then the value will be added to the current value otherwise a pair consisting of $j$ and its value is inserted  into the hash table.  
\begin{algorithm}
\caption{Row-wise algorithm for numeric calculation of $AP$. \label{alg:NumericAP}}
\begin{algorithmic}[1]
\STATE{Input: $A_l$, $P_l$, $\tilde{P}_{\text{r}}$}
\STATE{Initialize $R = \emptyset $}
\STATE Update $\tilde{P}_{\text{r}}$ using a sparse MPI communication
\STATE{$i=1$}
\FOR {$i<=$ the number of rows of $C_l$}
\STATE $R$ = Alg.~\ref{alg:NumericOneRowOfAP}($i$, $A_l$, $P_l$, $\tilde{P}_{\text{r}}$)
\STATE  Add  $R$ into the $i$th row of $C_l$
\STATE  Clear $R$
\STATE $i += 1$
\ENDFOR 
\STATE{Output: $C_l$ filled with  numeric values}
\end{algorithmic}
\end{algorithm}
In Alg.~\ref{alg:NumericAP},  we extract keys and their corresponding values from $R$, and call the matrix function, {\it MatSetValues}, in PETSc directly  to add the values to the allocated space in  $C_l$. ``Clear" at line 8 of Alg.~\ref{alg:NumericAP} does not deallocate memory, instead, it resets a flag so that the memory can be reused for next row.

\section{All-at-once algorithms  for sparse matrix triple products}
In multigrid/multilevel  methods, to construct a scalable parallel  solver, coarse spaces are included, which can be built either geometrically (\cite{kong2016highly, kong2017scalable}) or algebraically (\cite{kong2018fully, kong2019highly}).  One of critical tasks is to form a coarse operator, $C$, based on the interpolation, $P$, and the fine level operator, $A$,  as follows
\begin{equation}\label{eq:tripleproduct}
C = P^TAP,
\end{equation}
where the notation $C$ is reused to denote  the output matrix of the sparse matrix triple product without confusion.  It is straightforward to apply Alg.~\ref{alg:symbolicAP} and~\ref{alg:NumericAP} twice for computing  the sparse matrix triple product, that is,
\begin{equation}\label{eq:tripleproductsplit}
\begin{array}{lll}
\tilde{C} = AP, \\
C = P^T \tilde{C},
\end{array}
\end{equation}
where $\tilde{C}$ is an auxiliary  matrix.  More precisely, $\tilde{C} = AP$ is explicitly implemented  using Alg.~\ref{alg:symbolicAP} for the symbolic calculation  and Alg.~\ref{alg:NumericAP} for the numeric computation, and then the second product is formed using  the same algorithms  again, $C=P^T \tilde{C}$.  The procedure (referred to as ``two-step" method) is  summarized in Alg.~\ref{alg:SymbolicPtAP} for the symbolic caculation and Alg.~\ref{alg:NumericPtAP} for the numeric calculation.  
\begin{algorithm}
\caption{Two-step method for symbolic calculation of $P^TAP$. \label{alg:SymbolicPtAP}}
\begin{algorithmic}[1]
\STATE{Input: $A_l$, $P_l$}
\STATE{$\tilde{C}_l$ = Alg.~\ref{alg:symbolicAP}($A_l$,$P_l$) }
\STATE{$P^T_l = [P^T_{\text{d}}, P^T_{\text{o}}] = $ Explicit  symbolic-transpose($P_l$})
\STATE Symbolically compute $C_{\text{s}} = P^T_{\text{o}}  A_l$
\STATE Send $C_{\text{s}} $ to its owners
\STATE Symbolically compute $C_l = P^T_{\text{d}}  A_l$
\STATE  Receive $C_{\text{r}}$ from remote processors
\STATE  $C_l += C_{\text{r}}$
\STATE{Output: $C_l$ with allocated space }
\end{algorithmic}
\end{algorithm}
The lines 4 and 6 of Alg.~\ref{alg:SymbolicPtAP} are computed using a similar algorithm as Alg.~\ref{alg:symbolicAP} but without extracting any remote rows of $\tilde{C}_l$.  The communication at the lines 5 and 7 is implemented  using  nonblocking MPI techniques so that the computation and the communication are overlapped. ``+=" at the line 8 of Alg.~\ref{alg:SymbolicPtAP} represents adding data from $C_{\text{r}}$ to $C_l$. 
\begin{algorithm}
\caption{Two-step method for numeric calculation of $P^TAP$. \label{alg:NumericPtAP}}
\begin{algorithmic}[1]
\STATE{Input: $A_l$, $P_l$}
\STATE{$\tilde{C}_l$ = Alg.~\ref{alg:NumericAP}($A_l$,$P_l$) }
\STATE{$P^T_l= [P^T_{\text{d}}, P^T_{\text{o}}] $ = Numeric-transpose($P_l$)}
\STATE Numerically  compute $C_{\text{s}} = P^T_{\text{o}}  A_l$
\STATE Send $C_{\text{s}} $ to its owners
\STATE Numerically compute $C_l = P^T_{\text{d}}  A_l$
\STATE  Receive $C_{\text{r}}$ from remote processors
\STATE  $C_l += C_{\text{r}}$
\STATE{Output: $C_l$ with filled numeric  values}
\end{algorithmic}
\end{algorithm}
Similarly, the lines 4 and 6 of Alg.~\ref{alg:NumericPtAP} are implemented using a similar algorithm as Alg.~\ref{alg:NumericAP} without extracting any remote rows of $\tilde{C}_l$. ``+=" operator at the line 8 is implemented using {\it MatSetValues} in PETSc that efficiently adds values to the diagonal and off-diagonal matrices. The advantage of the two-step  algorithm is that it can be efficiently implemented using the row-wise algorithm. However, this algorithm needs to build auxiliary matrices  such as $P^T$ (explicit transpose of $P$) and $\tilde{C}$, which leads to a high  memory overhead. The two-step algorithm works  well for some applications  that do not require much memory, but it is not suitable for memory-intensive   problems such as seven-dimensional  neutron transport simulations. To overcome the difficulty, in this work, we introduce  new all-at-once algorithms that do  not involve  the auxiliary matrices  for saving memory. The basic idea of the all-at-once algorithms  is to form  $C$ with one pass through $P^T$, $A$ and $P$ all at once without any auxiliary  steps.  We also want to mention that the similar idea has been also used in (\cite{adams2004ultrascalable}, \cite{yang2002boomeramg}). However, we propose a new version  of the all-at-once algorithms  based on hash tables that is implemented in PETSc as part of this work, where the second matrix-matrix multiplication is based on an outer product algorithm instead of the row-wise approach. More precisely, the $j$th  element of the $i$th row of $C$ is obtained using the following formula
\begin{equation} \label{eq:allatonce}
\begin{array}{llll}
C(i,j) &= \displaystyle \sum_{I} P^T(i,I)   \sum_{J} A(I,J) P(J,j), \\
& = \displaystyle  \sum_{I}  P(I,i)  \sum_{J} A(I,J) P(J,j),
\end{array}
\end{equation}
where we do not need to explicitly form $P^T$. The atomic task of the all-at-once algorithms  is to compute the $i$th row of $C$ as follows 
\begin{equation} \label{eq:allatoncerow}
C(i,:)  =  \displaystyle   \sum_{I}P(I,i)  \sum_{J} A(I,J) P(J,:). 
\end{equation}
Eq.~\eqref{eq:allatoncerow} potentially is memory efficient, but there is no a good way to access $P$ column by column since $P$ is distributed  in a row-wise  manner.   
To overcome this difficulty,  we use an outer  product for the second matrix-matrix multiplication, $P^T (AP)$, to  access $P$ row by row instead of column by column.   $C$ is formed all at once with a summation of outer  products,
\begin{equation} \label{eq:allatonceoutproduct}
C = \sum_{I} P(I,:)   \otimes \sum_{J} A(I,J) P(J,:). 
\end{equation}
In Eq.~\eqref{eq:allatonceoutproduct}, $C$ is the summation of the outer  product of  the $I$th row of $P$ and the $I$th row of $AP$. Note that (\cite{ballard2016reducing}) shows that, in their two-step approach, the row-wise algorithm is suitable for $AP$, and the outer  product is the best for $P^T (AP)$ in terms of the communication cost. We adopt the outer  product in the second matrix-matrix multiplicaiton not only for reducing communication cost but also for saving memory.  The detailed algorithm  is shown in Alg.~\ref{alg:SymbolicPtAPallatonce} and~\ref{alg:NumericPtAPallatonce}.
\begin{algorithm}
\caption{All-at-once algorithm for symbolic calculation of $P^TAP$. \label{alg:SymbolicPtAPallatonce}}
\begin{algorithmic}[1]
\STATE{Input: $A_l=[A_{\text{d}}, A_{\text{o}}]$, $P_l=[P_\text{d}, P_{\text{o}}]$}
\STATE {Extract the remote rows, $\tilde{P}_{\text{r}}$, of $P$  corresponding to the nonzero columns of $A_{\text{o}}$}
\STATE Initialize $C_{\text{s}}^H$ as a set of hash tables (each hash table corresponds to a row)
\STATE{$I=1$}
\FOR {$I<=$ the number of  rows of $A_l$}
\IF{$P_{\text{o}}(I, :)$ is empty}
\STATE $I+=1$
\STATE continue
\ENDIF
\STATE $[R_\text{d}, R_\text{o}]$ = Alg.~\ref{alg:onerowofAP}($I$, $A_l$, $P_l$, $\tilde{P}_{\text{r}}$) 
\STATE Symbolic  calculation: $C_{\text{s}}^H += P_{\text{o}}(I, :) \otimes [R_\text{d}, R_\text{o}]$ 
\STATE  $I+= 1$
\ENDFOR
\STATE Send $C_{\text{s}}^H$ to its owners
\STATE Initialize $C_{\text{l}}^H$ as a set of hash tables (two hash tables are needed for each row; one for the diagonal matrix and the other for the off-diagonal  matrix)
\STATE{$I=1$}
\FOR {$I<=$ the number of  rows of $A_l$}
\IF{$P_{\text{d}}(I, :)$ is empty}
\STATE $I+=1$
\STATE continue
\ENDIF
\STATE $[R_\text{d}, R_\text{o}]$ = Alg.~\ref{alg:onerowofAP}($I$, $A_l$, $P_l$, $\tilde{P}_{\text{r}}$)
\STATE Symbolic  calculation $C_{l}^H += P_{\text{d}}(I, :) \otimes [R_\text{d}, R_\text{o}]$ 
\STATE  $I += 1$
\ENDFOR
\STATE Receive $C_{\text{r}}^H$ from the  remote contributors 
\STATE Symbolic calculation: $C_l^H +=  C_{\text{r}}^H$
\STATE Free the memory of $C_{\text{r}}^H$
\STATE Initialize $\text{nzd}=\{0\}$ and $\text{nzo}=\{0\}$
\STATE $i=1$
\FOR {$i<=$ the number of rows of $C_l$}
\STATE$\text{nzd}(i) = |C_{\text{d}}^H(i,:)|  $
\STATE$\text{nzo}(i) = |C_{\text{o}}^H(i,:)|  $
\ENDFOR 
\STATE Free the memory of $C_l^H$
\STATE{ Preallocate memory for $C_l$ using $\text{nzd}$  and  ${\text{nzo}}$}
\STATE{Output: $C_l$ with allocated space}
\end{algorithmic}
\end{algorithm}
At the line 3 of Alg.~\ref{alg:SymbolicPtAPallatonce},  $C_{\text{s}}^H$ is implemented as a set of hash sets that stores the indices of  the nonzero columns for each row.  The outer product at the line 11 is carried out by inserting the $I$th row of $AP$, $R=[R_d, R_o]$,  into the rows of $C_{\text{s}}^H$ corresponding to the nonzero columns of $P_{\text{o}}(I, :)$.  The first loop consisting of  lines 5-13 computes the remote rows that are sent to its remote  owners.  The second loop consisting of  lines 16-25 calculates the local part of $C$.  The calculation is split into two parts for overlapping  the computation and the communication.
\begin{algorithm}
\caption{ All-at-once algorithm for numeric calculation of $P^TAP$. \label{alg:NumericPtAPallatonce}}
\begin{algorithmic}[1]
\STATE{Input: $A_l=[A_{\text{d}}, A_{\text{o}}]$, $P_l=[P_\text{d}, P_{\text{o}}]$}
\STATE {Extract the remote rows, $\tilde{P}_{\text{r}}$, of $P$  corresponding to the nonzero columns of $A_{\text{o}}$}
\STATE{$I=1$}
\FOR {$I<=$ the number of  rows of $A_l$}
\IF{$P_{\text{o}}(I, :)$ is empty}
\STATE $I+=1$
\STATE continue
\ENDIF
\STATE $R$ = Alg.~\ref{alg:NumericOneRowOfAP}($I$, $A_l$, $P_l$, $\tilde{P}_{\text{r}}$)
\STATE Numeric calculation: $C_{\text{s}} += P_{\text{o}}(I, :) \otimes R$ 
\STATE  $I += 1$
\ENDFOR
\STATE Send $C_{\text{s}}$ to its owners
\STATE{$I=1$}
\FOR {$I<=$ the number of  rows of $A_l$}
\IF{$P_{\text{d}}(I, :)$ is empty}
\STATE $I+=1$
\STATE continue
\ENDIF
\STATE $R$ = Alg.~\ref{alg:NumericOneRowOfAP}($I$, $A_l$, $P_l$, $\tilde{P}_{\text{r}}$)
\STATE Numeric calculation: $C_{l} += P_{\text{d}}(I, :) \otimes R$ 
\STATE  $I += 1$
\ENDFOR
\STATE Receive $C_{\text{r}}$ from remote contributors 
\STATE $C_l += C_{\text{r}}$
\STATE Free the memory of $C_{\text{r}}$
\STATE{Output: $C_l$} filled with numeric values
\end{algorithmic}
\end{algorithm}
In Alg.~\ref{alg:NumericPtAPallatonce},  adding  the outer  product of $P_{\text{o}}(I, :) \otimes R$ at the line 10 is implemented by inserting the numerical values directly into the preallocated space,
and the outer  product at the line 21 is added to the preallocated matrix using PETSc routine {\it MatSetValues}. Again, the communications at the lines 14 and 26 of  Alg.~\ref{alg:SymbolicPtAPallatonce} are based on nonblocking MPI techniques.  It is obviously  observed  that  Alg.~\ref{alg:SymbolicPtAPallatonce} and~\ref{alg:NumericPtAPallatonce} do not involve auxiliary  matrices $P^T$ and $\tilde{C}$.  Generally,  Alg.~\ref{alg:SymbolicPtAPallatonce} and~\ref{alg:NumericPtAPallatonce} work great when the amount of the calculations in the first loop is small (it is true for most of the sparse matrices arising from the discretization of partial differential  equations). If the computation of the first loop is heavy, the algorithm can be modified to save  the time on the  calculations. In this case, we propose  an alternative choice that merges the calculations  in the first loop and that in  the second loop together. This modified algorithm is referred to as ``merged all-at-once".  The merged all-at-once algorithm is described in detail  in Alg.~\ref{alg:SymbolicPtAPallatonceMerged}. and~\ref{alg:NumericPtAPallatonceMerged}.
\begin{algorithm}
\caption{Merged all-at-once algorithm for symbolic calculation of $P^TAP$. \label{alg:SymbolicPtAPallatonceMerged}}
\begin{algorithmic}[1]
\STATE{Input: $A_l=[A_{\text{d}}, A_{\text{o}}]$, $P_l=[P_\text{d}, P_{\text{o}}]$}
\STATE {Extract the remote rows, $\tilde{P}_{\text{r}}$, of $P$  corresponding to the nonzero columns of $A_{\text{o}}$}
\STATE Initialize $C_{\text{s}}^H$ as a set of hash tables
\STATE Initialize $C_{l}^H$ as a set of hash tables
\STATE{$I=1$}
\FOR {$I<=$ the number of  rows of $A_l$}
\IF{both $P_{\text{o}}(I, :)$  and $P_{\text{d}}(I, :)$ are empty}
\STATE $I+=1$
\STATE continue
\ENDIF
\STATE $[R_\text{d}, R_\text{o}]$ = Alg.~\ref{alg:onerowofAP}($I$, $A_l$, $P_l$, $\tilde{P}_{\text{r}}$)
\STATE Symbolic  calculation $C_{\text{s}}^H += P_{\text{o}}(I, :) \otimes [R_\text{d}, R_\text{o}]$ 
\STATE Symbolic  calculation $C_{l}^H += P_{\text{d}}(I, :) \otimes [R_\text{d}, R_\text{o}]$ 
\STATE  $I+= 1$
\ENDFOR
\STATE Send $C_{\text{s}}^H$ to its owners
\STATE Receive $C_{\text{r}}^H$ from the  remote contributors 
\STATE Symbolic calculation $C_l^H  += C_{\text{r}}^H$
\STATE Free the memory of $C_{\text{r}}^H$
\STATE Initialize $\text{nzd}=\{0\}$ and $\text{nzo}=\{0\}$
\STATE $i=1$
\FOR {$i<=$ the number of rows of $C_l$}
\STATE$\text{nzd}(i) = |C_{\text{d}}^H(i,:)|  $
\STATE$\text{nzo}(i) = |C_{\text{o}}^H(i,:)|  $
\ENDFOR 
\STATE Free the memory of $C_l^H$
\STATE{ Preallocate memory for $C_l$ using $\text{nzd}$  and  ${\text{nzo}}$}
\STATE{Output: $C_l$ with allocated memory}
\end{algorithmic}
\end{algorithm}
\begin{algorithm}
\caption{Merged all-at-once algorithm for numeric calculation of $P^TAP$. \label{alg:NumericPtAPallatonceMerged}}
\begin{algorithmic}[1]
\STATE{Input: $A_l=[A_{\text{d}}, A_{\text{o}}]$, $P_l=[P_\text{d}, P_{\text{o}}]$}
\STATE {Extract the remote rows, $\tilde{P}_{\text{r}}$, of $P$  corresponding to the nonzero columns of $A_{\text{o}}$}
\STATE{$I=1$}
\FOR {$I<=$ the number of  rows of $A_l$}
\IF{both $P_{\text{o}}(I, :)$  and  $P_{\text{d}}(I, :)$ are empty}
\STATE $I+=1$
\STATE continue
\ENDIF
\STATE $R$ = Alg.~\ref{alg:NumericOneRowOfAP}($I$, $A_l$, $P_l$, $\tilde{P}_{\text{r}}$)
\STATE Numeric calculation $C_{\text{s}} += P_{\text{o}}(I, :) \otimes R$ 
\STATE Numeric calculation: $C_{l} += P_{\text{d}}(I, :) \otimes R$ 
\STATE  $I += 1$
\ENDFOR
\STATE Send $C_{\text{s}}$ to its owners
\STATE Receive $C_{\text{r}}$ from the remote contributors 
\STATE $C_l  += C_{\text{r}}$
\STATE Free the memory of $C_{\text{r}}$
\STATE{Output: $C_l$} filled with numeric values
\end{algorithmic}
\end{algorithm}
The performance differences between the all-at-once algorithm and its merged version are totally problem dependent. For some applications, the performances may be close to each other.
If the commutation in the first loop   is expensive, we may prefer the all-at-once to  the merged all-at-once.

\section{Numerical results}
In this section, we report the performance of the proposed algorithms in terms of the memory usage and the compute time. For understanding  the algorithms, we study two tests cases; a model problem based on structured meshes for mimicking  geometric muligrid/multilevel  methods, and a realistic neutron transport simulation in which  AMG is employed.  Let us define some notations that are used in the rest of discussions  for convenience.  ``two-step" represents the approach described in  Alg.~\ref{alg:SymbolicPtAP} and~~\ref{alg:NumericPtAP}. ``allatonce"  is the agorithm described in Alg.~\ref{alg:SymbolicPtAPallatonce} and~\ref{alg:NumericPtAPallatonce}. ``merged" denotes the method in Alg.~\ref{alg:SymbolicPtAPallatonceMerged} and~\ref{alg:NumericPtAPallatonceMerged}. ``$np$" is the number of processor cores, ``Mem" is the estimated memory usage per processor core,  in Megabyte, for the matrix triple products,  ``Mem$_{\text{T}}$" is the estimated  total memory per processor core, in Megabyte, used in the entire simulation, ``Time" is the compute time, in second, of the matrix triple products, ``Time$_{\text{T}}$" is the total compute time, in second, in the simulation, 
``Time$_{\text{sym}}$" is the time spent on the symbolic calculations of the triple products, ``Time$_{\text{num}}$" is the time on the numeric computations of the triple products,  and ``EFF" is the parallel efficiency of the overall simulation.  The proposed algorithms, all-at-once and its merged version, for the matrix triple products  are implemented in PETSc (\cite{petsc-user-ref}) as part of this work. 

\subsection{Model problem}
\begin{figure}
 \centering
 \includegraphics[width=1\linewidth]{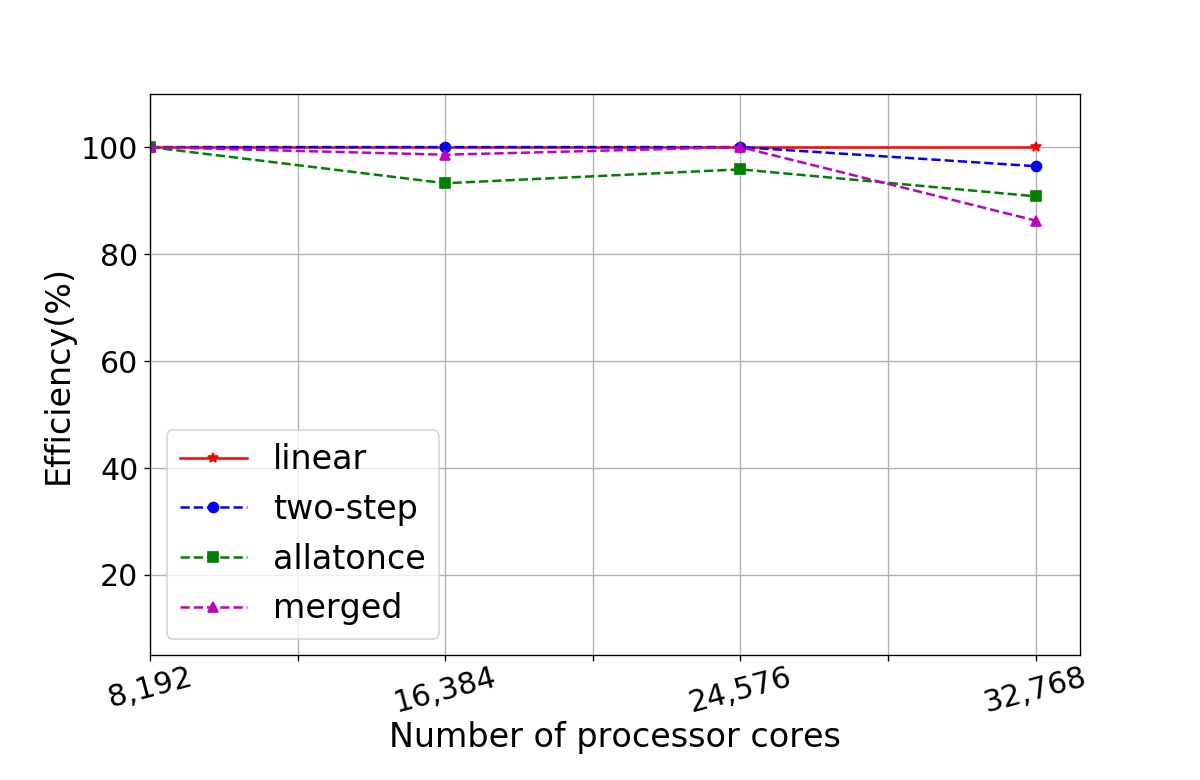} 
 \includegraphics[width=1\linewidth]{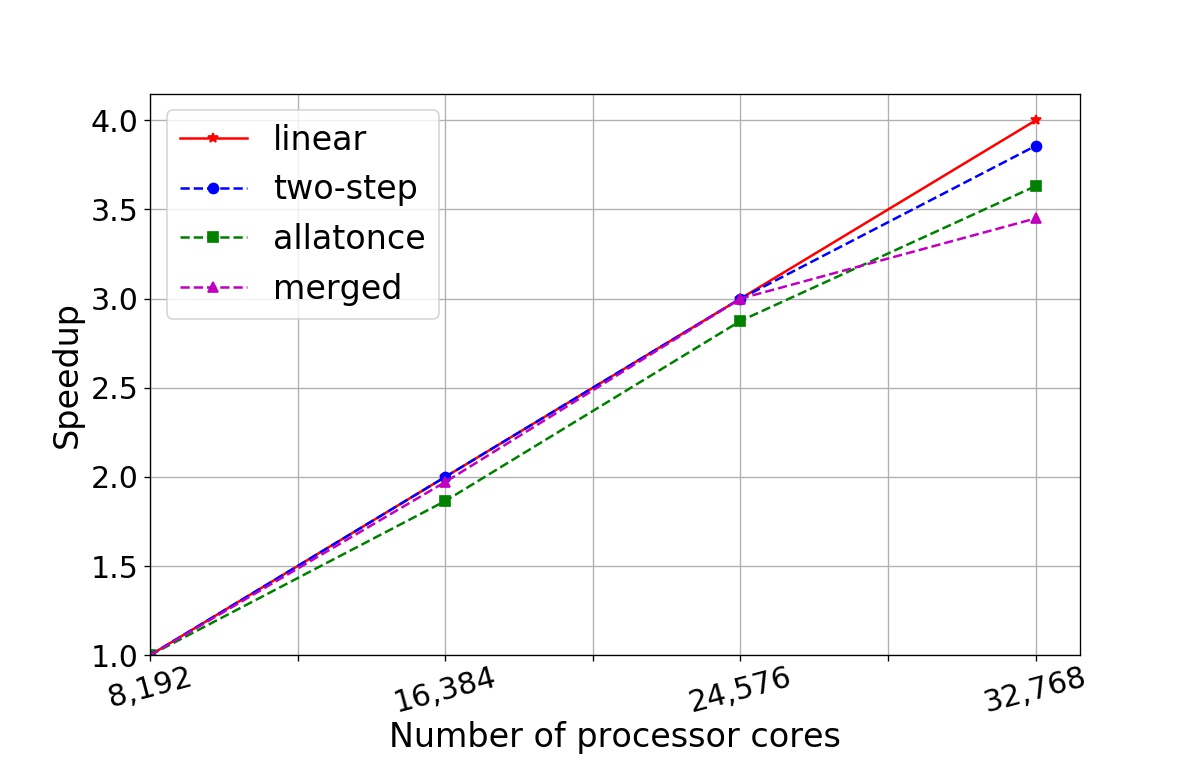} 
 \caption{Speedups and parallel efficiencies  of  different triple product algorithms on a structured mesh model problem  with $7,988,005,999$ unknowns  on $8,192$, $16,384$, $24,576$ and $32,768$ processor cores. Top: speedups, bottom: parallel efficiencies.\label{fig:structuredmeshspeedups}}
\end{figure}

This test is conducted for processor counts between 8,192 and 32,768 on the Theta supercomputer at Argonne National Laboratory. Theta is a massively parallel, many-core system with
second-generation Intel Xeon Phi processors. Each compute node has a 64-core processor with 16 gigabytes (GB) of high-bandwidth in-package memory (MCDRAM), 192 GB of DDR4 RAM, and a 128 GB SSD.  The  test is designed to  mimic a geometric  two-level  method based on a fine mesh and a coarse mesh.  A $1,000 \times 1,000 \times 1,000$ 3D structured grid  is  employed as the course mesh, and the fine mesh is an uniform refinement of the coarse mesh. Each grid point is assigned with one unknown.   An interpolation is created from the coarse mesh to the fine mesh using a linear function.  The dimensions of $A$ on the fine mesh are $7,988,005,999 \times 7,988,005,999$, and these of $P$ from the coarse mesh to the fine mesh  are $7,988,005,999  \times 1,000,000,000$. One symbolic  and eleven  numeric triple products are implemented to mimic the use case in which  the numeric calculations  need to be repeated multiple times.
``Time$_{\text{num}}$" is the time on all  eleven numeric triple products.
 We compare  the performance of our new algorithms  in terms of the compute time and the memory usage with that obtained using the traditional two-step method. Numerical results are summarized in Table~\ref{tab:comparisonAllatonce_model}.
\begin{table}
\scriptsize
\centering
\caption{Memory usages and compute times of different triple product algorithms for a structured mesh model  problem with $7,988,005,999$ unknowns  on $8,192$, $16,384$, $24,576$ and $32,768$ processor cores. \label{tab:comparisonAllatonce_model}}
\begin{tabular}{c c c c c c  c c c c}
\toprule
$np$  & Algorithm&Mem & Time$_{\text{sym}}$ & Time$_{\text{num}}$&Time& EFF \\
\midrule
8,192 &allatonce  &  68 &  6.4 & 63&69&  100\%  \\
8,192 & merged   &  68 &6.3 &63&69 & 100\% \\
8,192 & two-step  &  554 &8.3 &46&54& 100\% \\
\midrule
16,384 & allatonce &  35 &  3.4 & 33&37&  93\%  \\
16,384 & merged  &  35 &3.2 &32&35 & 99\% \\
16,384& two-step  &  280 &4.2 &23&27& 100\% \\
\bottomrule
24,576 & allatonce  &  24 &  2.3 & 22&24&  96\%  \\
24,576 & merged  &  24 &2.2 &21&23 & 100\% \\
24,576 &  two-step  &  188 &2.9 &15&18& 100\% \\
\bottomrule
32,768 & allatonce  &  18 &  1.8 & 17&19&  91\%  \\
32,768 & merged  &  18 &1.7 &18&20 & 86\% \\
32,768 &  two-step  &  132 &2.2 &12&14& 96\% \\
\bottomrule
\end{tabular}
\end{table}
We observed that the memory usage  of the new algorithms is roughly  $10\%$ of  that consumed  by  the two-step method  from Table~\ref{tab:comparisonAllatonce_model}, which indicates that  the new algorithms produce a very low  memory overhead.   For example,  the all-at-once algorithm
takes only $68$ M memory at 8,192 processor cores, while the two-step method uses $554$ M. For all cases, the all-at-once  and the merged all-at-once approaches use exactly the same amount of memory.  All the algorithms are scalable in terms of the memory usage in 
the sense that the memory usages are proportionally decreased when the number of processor cores is increased. The memory usages are halved to $35$ M for 
the  all-at-once algorithm and $280$ M for the two-step method, respectively,  when the number of processor cores is doubled  from $8,192$ to $16,384$. They continue
being reduced to $18$ M and $132$ M, respectively,  when we increase the number of processor cores to $32,768$. 
The memory (denoted as ``Mem" in Table~\ref{tab:comparisonAllatonce_model}) spent on the triple products includes the storage for the output matrix $C$.   In order to understand how much memory overhead we have on the triple product 
algorithms, the memories used to store matrices $A$, $P$ and $C$ using  $8,192$, $16,384$, $24,576$ and $32,768$ processor cores are shown in Table~\ref{tab:matrixmemory}.  It is easily found 
that $C$ takes $65$ M at $8,912$ processor cores, and then the overhead of the all-at-once algorithm is $3$ M since the corresponding  total memory on the triple product for the all-at-once approach    is $68$ M. On the other hand, 
the two-step method has a much higher memory overhead, $489$ M ($554$ M - $65$ M). The memory usages  for storing all the matrices are scalable, that is,  they are ideally halved when we double the number of processor  cores.
For example, the memory required to store  $A$  is $182$ M at $8,192$ processor cores, and it is reduced to $92$ M when we increase the number of processor cores from 
$8,192$ to $16,384$. It continues being reduced to $62$ M and $47$ M, when we increase the number of processor cores to $24,576$ and $32,768$.  We observed  that the same trend in the memory usages  for storing $P$ and $C$, that is, they are perfectly halved when the number of processor cores is doubled.     Similarly, the compute times on the triple products  are also perfectly scalable for all the algorithms. The compute times using $8,192$ are $69$ s for the all-at-once method
and $54$ s for the two-step method, and they are  reduced to $37$ s and $27$ s when $16,384$ processor cores are used. This leads to parallel efficiencies of $93\%$ for the all-at-once
method and $100\%$ for the two-step method.  When we continue increasing the number of processor cores to $24,576$ and $32,768$, the compute times are reduced to $24$ s and $19$ s
for the all-at-once algorithm, and $18$ s and $14$ s for the two-step method.  A perfect scalability with  parallel efficiencies of above $90\%$ for all the algorithms is achieved even when the number of processor cores 
is up to $32,768$.  The compute time of the all-at-once approach  is very close to that using the merged  all-at-once algorithm, and there are no meaningful differences.  The two-step method is slightly faster than the all-at-once and the merged all-at-once algorithms.  These speedups and parallel efficiencies are also shown in Fig.~\ref{fig:structuredmeshspeedups}. 
\begin{figure}
 \centering
 \includegraphics[width=1\linewidth]{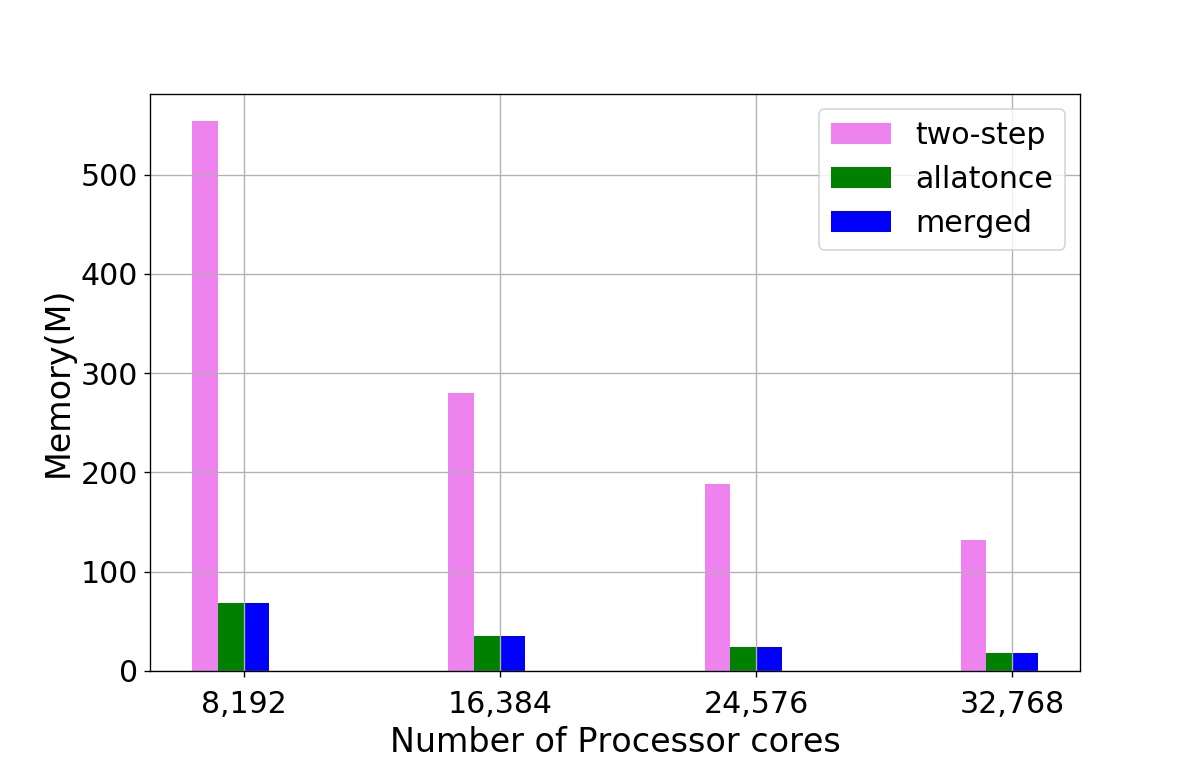} 
 \caption{Memory usages of different triple-product algorithms  on a structured mesh model problem  with $7,988,005,999$ unknowns  on $8,192$, $16,384$, $24,576$ and $32,768$ processor cores.  \label{fig:structuredmeshmemory}}
\end{figure}
\begin{table}
\scriptsize
\centering
\caption{Memories  for storing matrices $A$,  $P$ and $C$ for a structured mesh model problem  with $7,988,005,999$ unknowns using different numbers of processor cores; $8,192$, $16,384$, $24,576$ and $32,768$.\label{tab:matrixmemory}}
\begin{tabular}{c c c c c c  c c c c}
\toprule
Matrices  & 8,192&16,384 & 24,576 & 32,768 \\
\midrule
A &182 & 92 &  62 & 47 \\
P &125   &  63 &44 &26 \\
C  & 65 & 34 & 23 & 17 \\
\bottomrule
\end{tabular}
\end{table}
One more interesting thing is that the time spent on the symbolic calculation  for the all-at-once method is less than that spent by the two-step method. For  example, for the $8,192$-core 
case, $6.4$ s is used for the symbolic calculations in the all-at-once approach while the two-step method takes $8.3$ s. The times spent on both the symbolic and the numeric calculations
are ideally scalable.  In all, the new algorithms  are perfectly scalable in terms of the memory usage and the compute time, and they also use  much less memory than the traditional two-step method. The memory comparison  between all the algorithms are also summarized in Fig.~\ref{fig:structuredmeshmemory}, where it is easily found that the two-method approach consumes 
much more memory than the all-at-once and the merged all-at-once methods for all processor counts.

\begin{figure}
 \centering
 \includegraphics[width=1\linewidth]{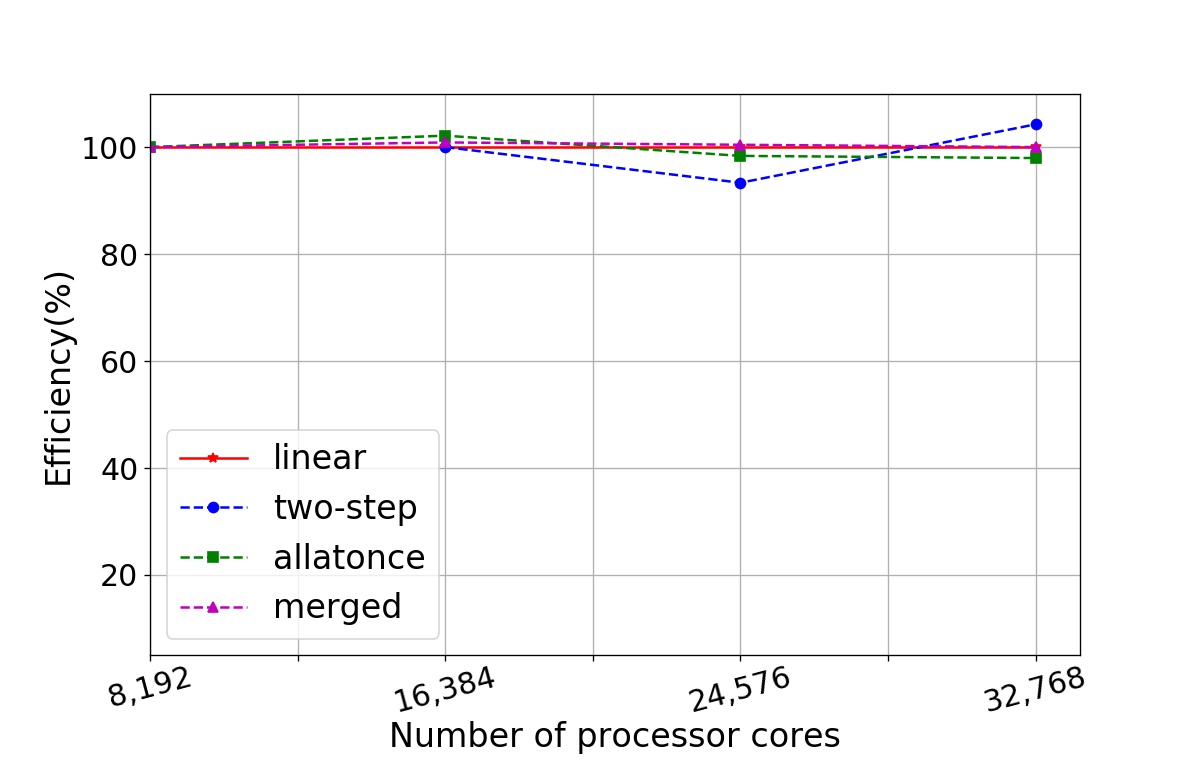} 
 \includegraphics[width=1\linewidth]{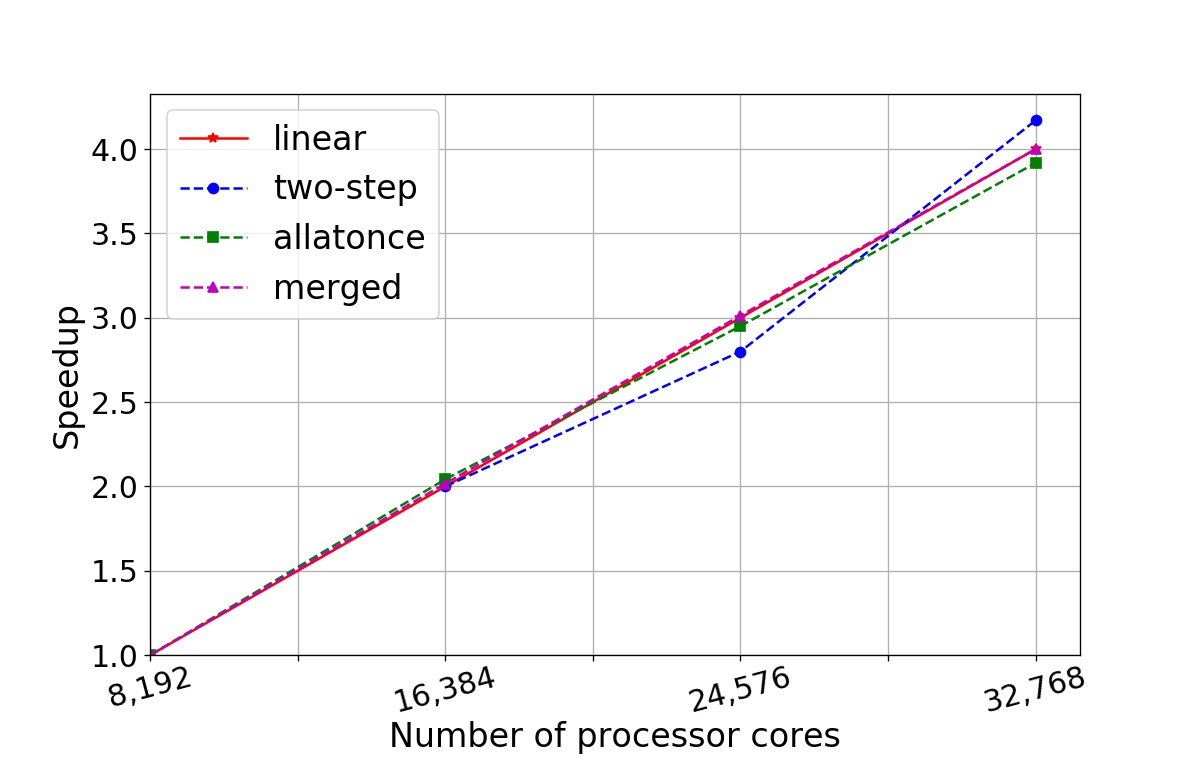} 
 \caption{Speedups and parallel efficiencies of different triple product algorithms for a 3D structured mesh model   problem with $26,973,008,999$ unknowns using $8,192$, $16,384$, $24,576$ and $32,768$ processor cores. Top: speedups, bottom: parallel efficiencies.\label{fig:structuredmeshfinespeedups}}
\end{figure}

\begin{figure}
 \centering
 \includegraphics[width=1\linewidth]{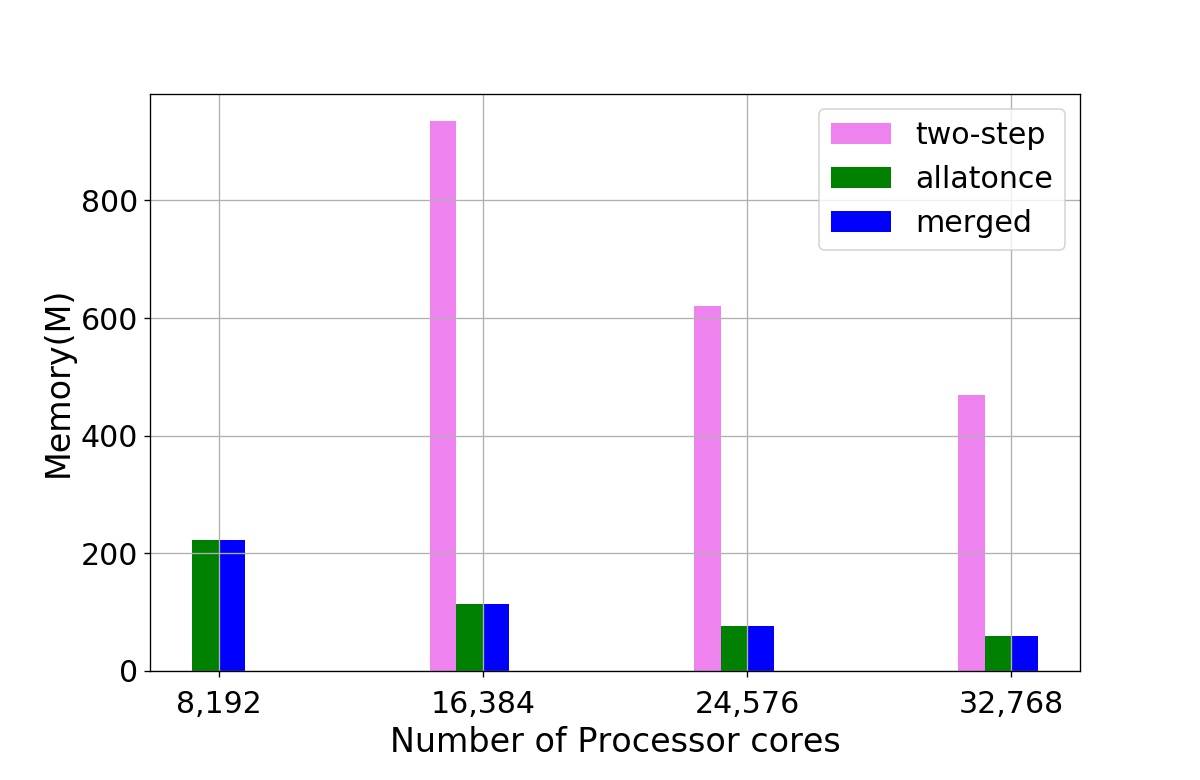} 
 \caption{Memory usages of different triple product algorithms for a 3D structured mesh model   problem with $26,973,008,999$ unknowns using $8,192$, $16,384$, $24,576$ and $32,768$ processor cores. \label{fig:structuredfinemeshmemory}}
\end{figure}

To further confirm the performance of the new algorithms,  larger meshes  are used in the following test. A $1500 \times 1500 \times 1500$ 3D structured grid is used as the coarse mesh, and the fine mesh   is an uniform  refinement of the coarse mesh. The dimensions of  $A$ on the fine mesh are  $26,973,008,999 \times 26,973,008,999$, and the dimensions of $P$  from the coarse mesh to the fine mesh are   $26,973,008,999 
\times 3,375,000,000$. The same configuration as before is adopted.  The numerical results are shown in Table~\ref{tab:comparisonAllatonce_model_fine}.  The memories  used to store matrices $A$, $P$ and $C$ are recored in Table~\ref{tab:matrixmemory_fine}.
\begin{table}
\scriptsize
\centering
\caption{Memory usages and compute times of different triple product algorithms for a structured mesh module problem with $26,973,008,999$ unknowns  on $8,192$, $16,384$, $24,576$ and $32,768$ processor cores. \label{tab:comparisonAllatonce_model_fine}}
\begin{tabular}{c c c c c c  c c c c}
\toprule
$np$  & Algorithm&Mem & Time$_{\text{sym}}$ & Time$_{\text{num}}$&Time& EFF \\
\midrule
8,192 &allatonce  &  223 &  21 & 218&239&  100\%  \\
8,192 & merged   &  223 &21 &211&232 & 100\% \\
8,192 & two-step  &  -- &-- &--&--& --\% \\
\midrule
16,384 & allatonce &  114 &  11 &106& 117&  102\%  \\
16,384 & merged  &  114 &10 &105&115 & 101\% \\
16,384& two-step  &  935 &21 &77&98& 100\% \\
\bottomrule
24,576 & allatonce  &  76 &  7.2 & 73&81&  98\%  \\
24,576 & merged  &  76 &7 &70&77 & 100\% \\
24,576 &  two-step  &  621 &12 &58&70& 93\% \\
\bottomrule
32,768 & allatonce  &  59 &  5.4 & 55&61&  98\%  \\
32,768 & merged  &  59 &5.3 &53&58 & 100\% \\
32,768 &  two-step  &  469 &8.8 &38&47& 104\% \\
\bottomrule
\end{tabular}
\end{table}
\begin{table}
\scriptsize
\centering
\caption{Memories for storing matrices $A$,  $P$ and $C$ for a structured mesh model problem  with $26,973,008,999$ unknowns using different numbers of processor cores; $8,192$, $16,384$, $24,576$ and $32,768$. \label{tab:matrixmemory_fine}}
\begin{tabular}{c c c c c c  c c c c}
\toprule
Matrices  &8,192&16,384 & 24,576 & 32,768 \\
\midrule
A &612  &  307 &  204 & 154 \\
P &426   &  275 &143 &107 \\
C & 212 & 106 & 74 & 54 \\
\bottomrule
\end{tabular}
\end{table}
The two-step method was not able to run using $8,192$ processor cores since it was attempting  to allocate too much memory beyond the physics memory.  
The parallel efficiencies for the two-step method are computed based on the compute time using $16,384$ processor cores. Again, the memory usage of the all-at-once algorithm 
is the same as that of the merged all-at-once algorithm. It is $223$ M at $8,192$ processors cores, and reduced to $114$ M, $76$ M and $59$ M when the number of processor cores
is increased to $16,384$, $24,576$ and $32,768$.  In fact, it is ideally halved  when we double the number of processor cores.   The two-step method consumes  nine times as much  memory as the  all-at-once algorithm. The detailed comparison of the memory usages for all the algorithms is drawn in Fig.~\ref{fig:structuredfinemeshmemory}.   It is obviously observed that 
the memory used in the two-step method is significantly more than that used in the all-at-once and the merged all-at-once algorithms. The times spent on the symbolic and the numeric calculations  are scalable for all the algorithm with up to $32,768$ processor cores.  For example, the all-at-once algorithm uses $21$ s for the symbolic calculations at $8,192$ processor cores, and the time is reduced to $11$ s, $7.2$ s, and $5.4$ s when we increase the number of processor cores to $16,384$, $24,576$ and $32,768$.  The time on the numeric calculations for the all-at-once algorithm 
at $8,192$ processor core is $218$ s, and it perfectly  is halved to $106$ s when we double the number of processor cores to $16,384$. When we continue increasing the number of processor cores  to $24,575$ and $32,768$, and the time of the numeric calculations is further reduced to $73$ s and $55$ s. For the two-step method, the time on the symbolic computations at $16,384$
processor core is $21$ s, and continues being reduced to $12$ s and $8.8$ s when we use $24,576$ and $32,768$ processor cores. For the numeric calculations of the two-step method,  the time is $77$ s at $16,384$, and it is reduced to $58$ s and $38$ s when the number of processor cores is increased to $24,576$ and $32,768$.  Again, the merged all-at-once algorithm has a similar performance in terms of the memory usage and the compute time as the all-at-once algorithm.  The all-at-once and the merged all-at-once algorithms are faster in the symbolic calculations than the two-step method, and are a little slower than the two-step method 
for the numeric calculations.  The overall  calculation  time is  scalable because both the symbolic and the numeric calculations are perfectly scalable.  The perfect scalabilities in terms of 
the compute time for all the  algorithms are obtained with all processor counts.  The corresponding speedups and parallel efficiencies are also drawn in Fig.~\ref{fig:structuredmeshfinespeedups}. 

\subsection{Neutron transport problem}
Next, we present the performance of the proposed algorithms  in  the realistic neutron transport simulations. 
The numerical experiments in this section  are carried out on a supercomputer at INL (Idaho National Laboratory), where each compute node has two 20-core processors with 2.4 GHz  and the compute nodes  are connected by an OmniPath  network.
\subsubsection{Problem description}
\begin{figure}
 \centering
 \includegraphics[width=0.8\linewidth]{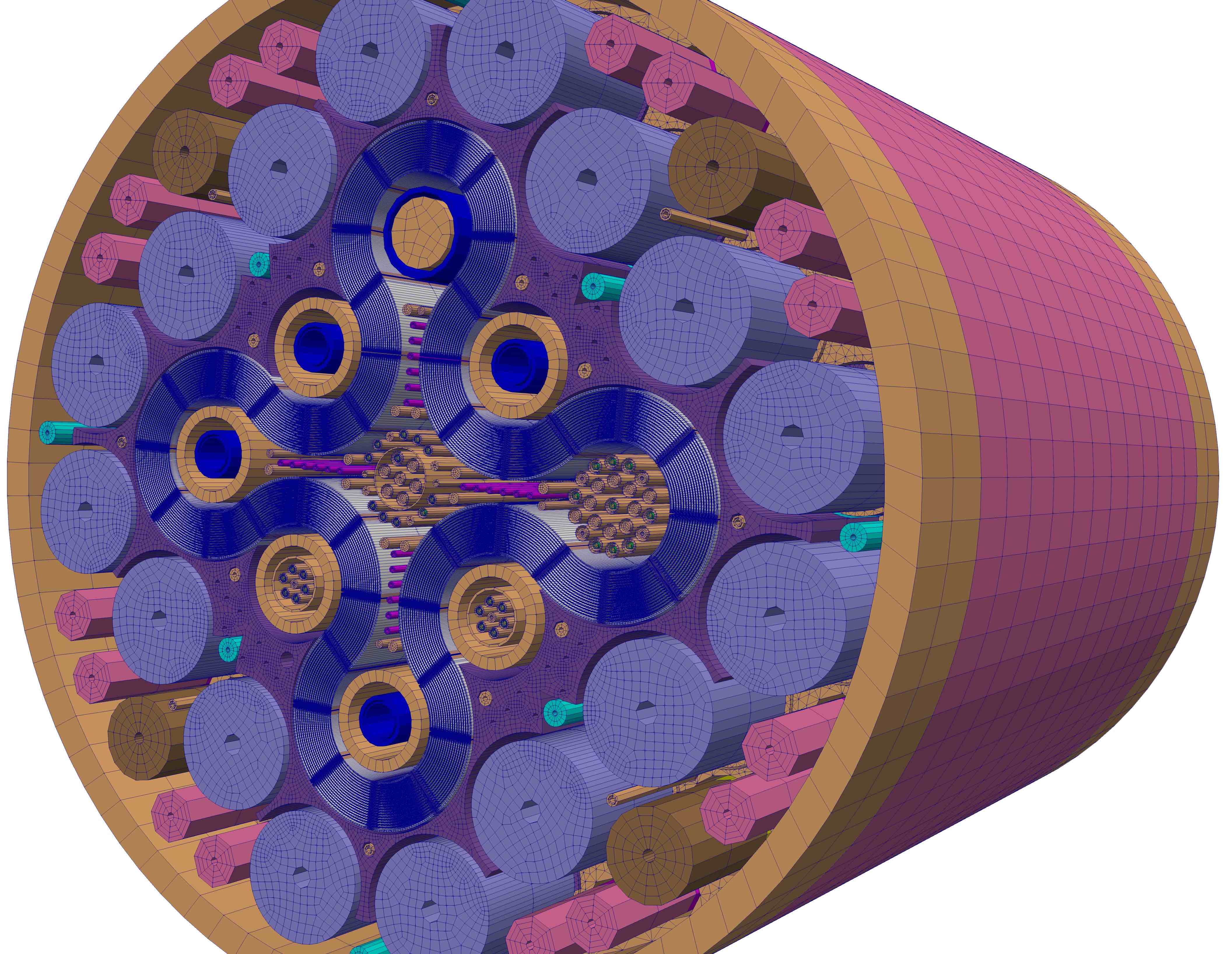} 
 \caption{Computational domain of ATR (Advanced Test Reactor at INL). \label{fig:ATRDomain}}
\end{figure}

We consider the multigroup neutron transport equations here, and  the neutron transport equations is a memory-intensive  application because there are  many variables  associated   with each  mesh vertex arising from the discretization  of the neutron flying direction. The  multigroup neutron transport  equations defined in $\domain \times \sphere$  ($\domain$ is the 3D spatial domain, e.g, shown in Fig.~\ref{fig:ATRDomain}, and $\sphere$ is the  2D unit sphere representing   the neutron flying directions) reads  as
\begin{subequations}
\footnotesize
\begin{equation}\label{eq:eigenvalue}
\begin{array}{llll}
\displaystyle
 \direction\cdot\grad\Psi_g +
 \Sigt[g]\Psi_g & =   \displaystyle
 \sum_{g'=1}^G  \displaystyle \int_{\sphere} \Sigs[g'\rightarrow g]  f_{g'\rightarrow g}(\direction'\cdot\direction)\Psi_{g'}(\position, \direction')\domg'  \\
 & \displaystyle + \frac{1}{4\pi}\frac{\chi_g}{k}\sum_{g'=1}^G\nu\Sigf[g']\Phi_{g'}, \text{ in } \domain \times \sphere, \\
\end{array}
 \end{equation} 
 \begin{equation}\label{eq:eigenvalue_boundary}
 \Psi_g =
 \alpha_g^{\text{s}}\Psi_g(\direction_r) \displaystyle+
 \alpha_g^{\text{d}}\frac{ \displaystyle \int_{\direction'\cdot\bnormal > 0} \left| \direction'\cdot\bnormal \right|\Psi_g \domg'}
 { \displaystyle \int_{\direction'\cdot\bnormal > 0} \left| \direction'\cdot\bnormal \right|\domg'},
 \text{ on }\partial\domain:\omgdnb < 0,
 \end{equation} 
\end{subequations}
\begin{figure}
 \centering
 \includegraphics[width=0.8\linewidth]{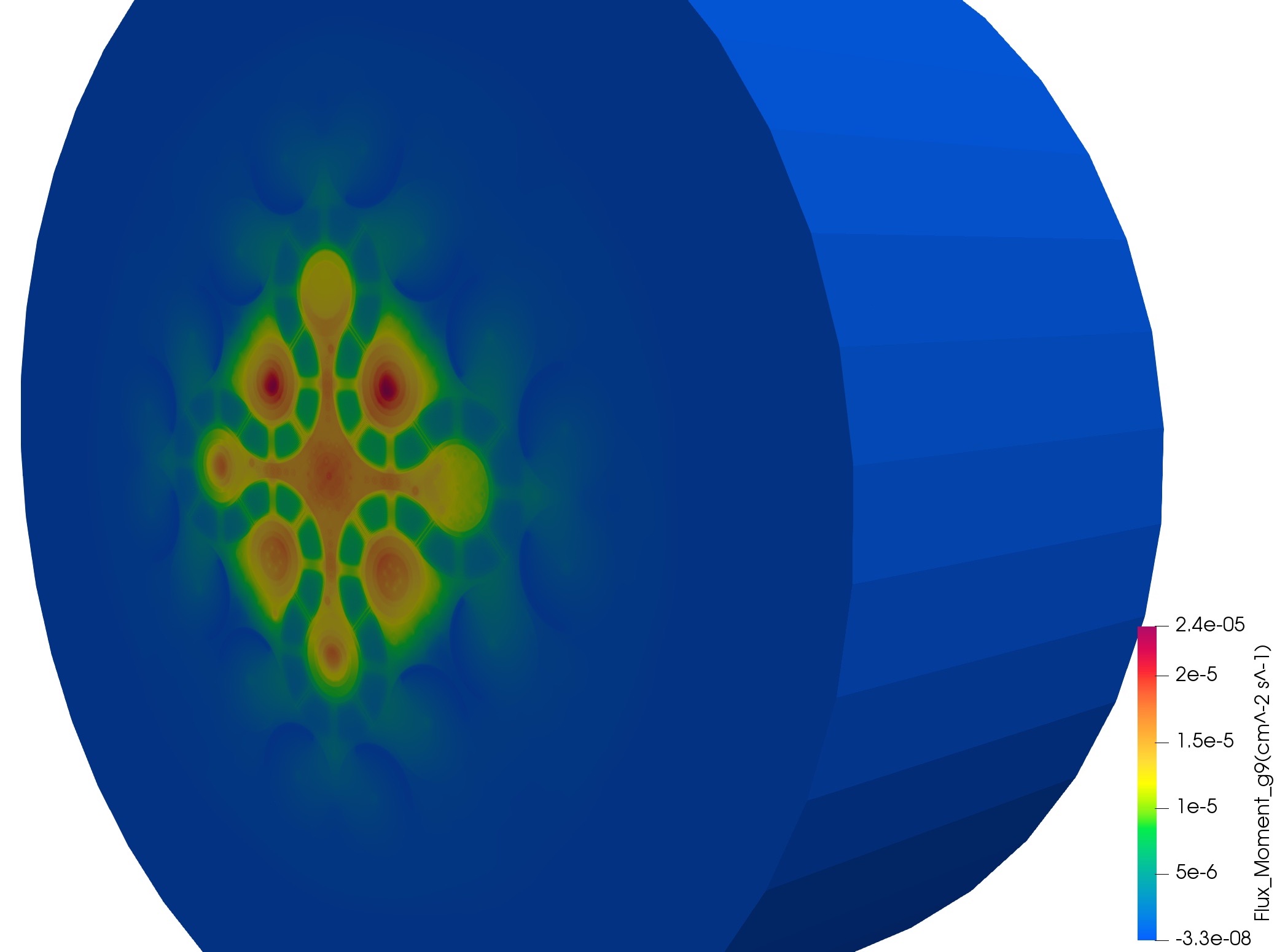} 
 \caption{9th flux moment. \label{fig:fluxmoment}}
\end{figure}
where  $g=1, 2,\cdots,G$, and  $G$ is the number of energy groups. The fundamental  quantity  of interest is neutron angular   flux $\Psi_g$  $[\afluxunit]$. In Eq.~\eqref{eq:eigenvalue}, $\direction \in \sphere$  denotes the independent angular variable,  $\position \in \domain$  is the independent spatial variable  $\left[ \unit{cm} \right]$,  $\Sigt[g]$ is the macroscopic total cross section $[ \unit{cm}^{-1} ]$,  $\Sigs[g'\rightarrow g]$ is the macroscopic scattering cross section from group $g'$ to group $g$  $[ \unit{cm}^{-1} ]$, $\Sigf[g]$ is the macroscopic fission cross section $[\unit{cm}^{-1}]$,  $\Phi_{g}$ is the scalar flux $[\sfluxunit ]$ defined as $\Phi_{g}  \equiv \int_\sphere \Psi_g  \domg$,  $f_{g'\rightarrow g}$ is the scattering phase function, $\nu$ is the averaged neutron emitted per fission,   $\chi_g$ is the prompt fission spectrum,   and $k$ is  the eigenvalue (sometimes referred to as a multiplication factor). In Eq.~\eqref{eq:eigenvalue_boundary}, $\bnormal$ is the outward unit normal vector on the boundary,  $\partial \domain$ is the boundary of  $\domain$,   $\direction_r = \direction - 2(\omgdnb)\bnormal$, $\alpha^\text{s}_g$ is the specular reflectivity on $\partial \domain$,  and $\alpha^\text{d}_g$ is the  diffusive reflectivity on $\partial \domain$.  The first term  of \eqref{eq:eigenvalue} is the \emph{streaming term}, and the second  is  the \emph{collision} term.  The first term of  the right hand side  of Eq.~\eqref{eq:eigenvalue}  is the \emph{scattering term}, which couples the angular fluxes of all directions and  energy  groups together. The second term of the right hand side of~\eqref{eq:eigenvalue} is the fission term,  which also couples the angular fluxes of all directions and 
 energy  groups together. Eq.~\eqref{eq:eigenvalue_boundary} is the boundary conditions.  The multigroup neutron transport equations are discretized  using  the first-order Lagrange finite element  in space, and the discrete ordinates method (that can be thought of as a collocation method) in angle (\cite{wang2018rattlesnake, lewis1984computational, kong2019highly}).  The discretizations are implemented based on RattleSnake (\cite{wang2018rattlesnake}), MOOSE (Multiphysics Object Oriented Simulation Environment) (\cite{peterson2018overview}) and libMesh (\cite{libMeshPaper}).  After  the angular and spatial discretizations,  we obtain a large-scale sparse eigenvalue system that is solved using an inexact  Newton-Krylov method  (\cite{cai1998parallel, knoll2004jacobian, kong2016parallel, kong2018efficient, kong2019highly}). During each Newton iteration, the Jacobian system is computed using GMRES (\cite{saad1986gmres}), and to speedup the convergence of GMRES, the multilevel  Schwarz preconditioner (\cite{kong2019highly}) is employed.  The triple products are performed in the setup phase of the multilevel  Schwarz preconditioner.  We next report the performance of the new all-at-once algorithms for the triple products within the multilevel method. 
 
 \subsubsection{Memory usage and scalability study}
\begin{figure}
 \centering
 \includegraphics[width=1\linewidth]{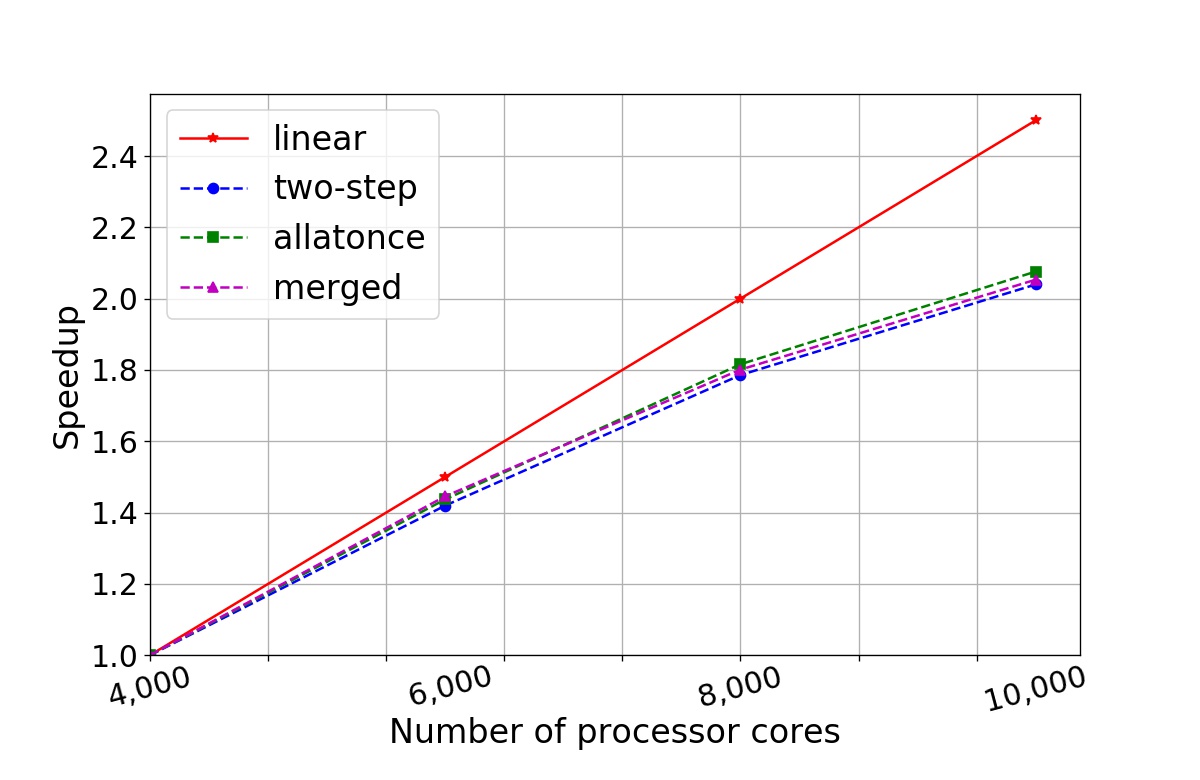} 
 \includegraphics[width=1\linewidth]{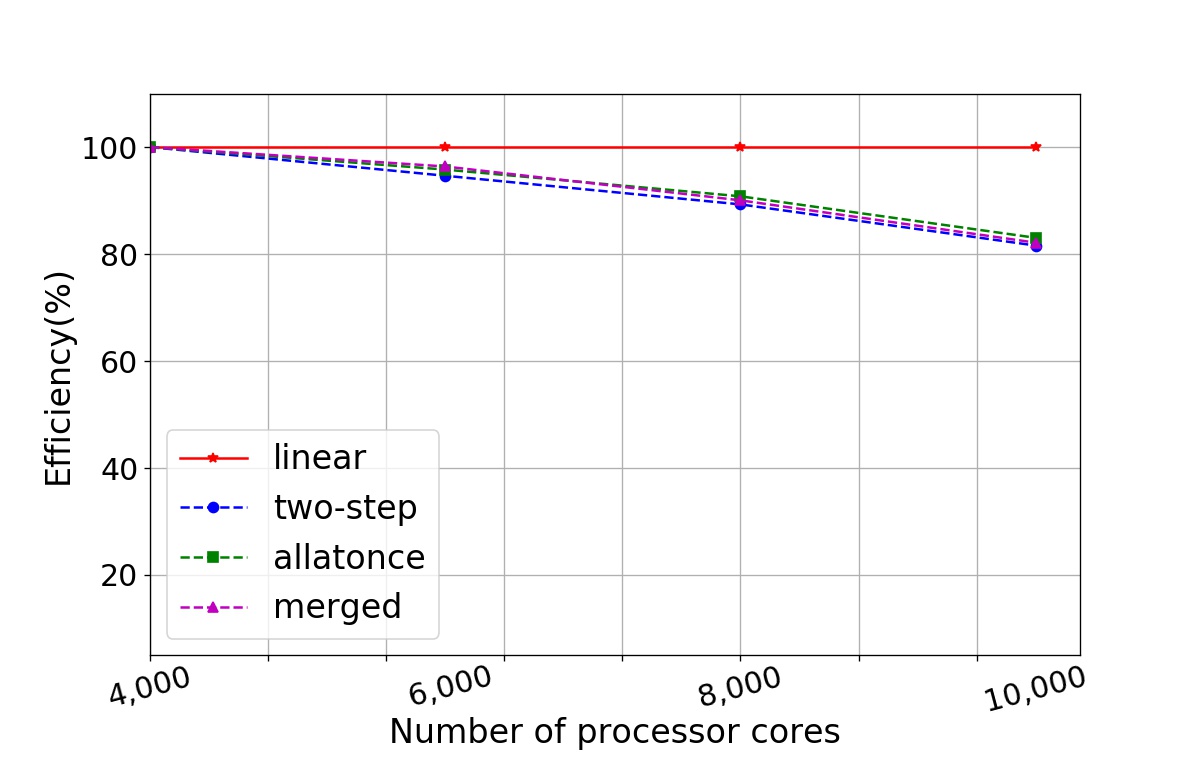} 
 \caption{Speedups and parallel efficiencies of the overall simulation using different triple-product algorithms without cacheing any intermediate data. Top: speedups, bottom: parallel efficiencies.\label{fig:s2speedups}}
\end{figure}

As below, we study the memory usages  of the all-at-once and the merged all-at-once algorithms compared with the two-step  approach.  In this test, a mesh with 25,856,505 nodes and 26,298,300 elements is employed, and 96 variables  are associated with each mesh vertex. The large-scale sparse eigenvalue system has 2,482,224,480 unknowns, which are computed with 4,000, 6,000, 8,000, and 10,000 processor cores, respectively. A flux moment of the physics solution is shown in Fig.~\ref{fig:fluxmoment}.  The preconditioning matrix is coarsened algebraically, and a twelve-level method is generated using 11 sparse matrix triple products, where $11$ interpolations and $12$ operator  matrices are formed. Interested  readers are referred to (\cite{kong2019highly})  for the coarsening algorithms  and the preconditioner setup.  The details of 12 operator matrices and 11 interpolation matrices for the twelve-level method  are shown in Table~\ref{tab:operators} and~\ref{tab:interpolations}.  Here ``level" is numbered from the fine level to the coarse levels, that is, ``level 0" is the finest level and ``level 11" is the coarsest level.  ``rows" is the number of the rows of  matrix, ``cols" is the number of the columns of matrix, ``nonzeros" denotes the number of nonzero elements of matrix, ``cols$_\text{min}$" is the minimum  number of nonzero columns among all rows, ``cols$_\text{max}$" is the maximum   number of nonzero columns among all rows, and ``cols$_\text{avg}$" is the averaged number of nonzero columns for all rows.  
\begin{table}
\scriptsize
\centering
\caption{ Operator matrices on different levels of the preconditioner.\label{tab:operators}}
\begin{tabular}{c c c c c c  c c c c}
\toprule
level  &rows &  nonzeros   &cols$_\text{min}$ &cols$_\text{max}$  & cols$_\text{avg}$\\
\midrule
0 & 2,482,224,480  &  66,169,152,672  &8&45&  26.7  \\
1 & 725,805,984         &  20,914,460,832     &3&101 & 28.8 \\
2 & 256,966,368        &  7,995,535,008       &1&172& 31.1 \\
3 &75,645,600           &  2,153,489,952        &4&278&  28.5  \\
4 & 21,120,960          &  617,242,944           &2&251 & 29.2 \\
5 & 5,187,840           &  169,280,256                   &3&288& 32.6 \\
6 &1,188,576            &  46,487,520                     &4&266&  39.1  \\
7 & 279,360             &  11,366,592                      &5&171 & 40.7 \\
8 & 63,648               &  2,076,192                           &4&86& 32.6 \\
9 & 13,632               &  319,680                        &7&51&  23.5  \\
10 & 2,496               &  37,824                              &8&25 & 15.2 \\
11 & 384                 &  1,536                                   &4&4& 4 \\
\bottomrule
\end{tabular}
\end{table}
\begin{table}
\scriptsize
\centering
\caption{Interpolation  matrices on different levels of the preconditioner.\label{tab:interpolations}}
\begin{tabular}{c c c  c c  c c c c}
\toprule
level  & rows&cols &cols$_\text{min}$  & cols$_\text{max}$  \\
\midrule
0 & 2,482,224,480  &  725,805,984 & 0  &12  \\
1 & 725,805,984  &  256,966,368   &0   &12 \\
2 & 256,966,368  &  75,645,600     &0   &11 \\
3 &75,645,600  &  21,120,960        &0   &9  \\
4 & 21,120,960  &  5,187,840         &0   &9 \\
5 & 5,187,840  &  1,188,576           &0   &8\\
6 &1,188,576 &  279,360               &0   &8 \\
7 & 279,360  &  63,648                 &0   &6 \\
8 & 63,648  &  13,632                   &0   &5\\
9 & 13,632  &  2,496                     &0   &4  \\
10 & 2,496  &  384                       &0   &2 \\
\bottomrule
\end{tabular}
\end{table}
The numerical results are summarized  in Table~\ref{tab:comparisonAllatonce} and~\ref{tab:comparisonAllatonceNofree}. It is easily observed, in Table~\ref{tab:comparisonAllatonce}, that the traditional two-step  algorithm uses twice as much  memory as  the all-at-once and the merged all-at-once algorithms do for all processor counts. For instance, the two-step  algorithm uses $587$ M memory at 4,000, while the all-at-once and the merged all-at-once algorithms use $264$ M memory  that is only $45\%$ of that used in the two-step method. This behavior is similar for all processor counts. The memory usages for all the algorithms are scalable when the number of processor cores is increased. They are $587$ M, $264$ M and  $264$ M  at 4,000 processor cores 
for the two-step, the all-at-once and the merged all-at-once algorithms, respectively, and proportionally  reduced to $310$ M, $158$ M, $158$ M when we increase the number of processor cores 
to 6,000. The memory usages continue being decreased to $283$ M,  $108$ M, and $108$ M, when the number of processor cores is increased to $8,000$. The memory usages 
at $10,000$ processor cores are slightly increased, but this does not affect the overall memory scalability. Meanwhile, the compute times of the all-at-once and the merged all-at-once approaches are almost the same as that spent using the two-step method. More important, the compute times for all three algorithms are aggressively optimized so that they account for  a negligible  portion of the total compute time.  The triple products do not affect the overall scalability and  performance of the simulations.  The compute times spent on the sparse matrix triple products sometimes  are  in the machine noise range  so that we observed the total compute time is slightly larger even when the compute time of the triple products is smaller. For example, the total compute time using the merged all-at-once algorithm  is $1977$ s at 6,000 processor cores  while that of the two-step method is $1990$ s even though the compute time of the triple product using the two-step method is smaller than that obtained using the merged all-at-once algorithm.  We therefore conclude that the proposed new algorithms including the all-at-once and the merged all-at-once approaches significantly   save  memory without any compromise on the compute time. The total compute time at 4,000 for the all-at-once algorithm is $2825$ s, and it is reduced to $1990$ s, 
$1582$ s, $1385$ s for 6,000, 8,000, 10,000 processor cores, respectively.  The compute times for the merged all-at-once and the two-step method algorithms are close to that of the all-at-once approach for all processor counts, that is, they are scalable as well. Good scalabilities and parallel efficiencies in terms of the compute time  for all the algorithms are maintained as shown in Fig.~\ref{fig:s2speedups}. A parallel efficiency of at least  $82\%$ is achieved at $10,000$ processor cores.
\begin{table}
\scriptsize
\centering
\caption{ Memory usages and compute times of different triple product algorithms without caching intermediate data on 4,000, 6,000, 8,000 and 10,000 processor cores.\label{tab:comparisonAllatonce}}
\begin{tabular}{c c c c c c  c c c c}
\toprule
$np$  & Algorithm&Mem & Mem$_T$ & Time & Time$_T$& EFF \\
\midrule
4,000 & two-step  &  587 &  1755 & 39&2825&  100\%  \\
4,000 & allatonce  &  264 &1423 &59&2871 & 100\% \\
4,000 & merged  &  264 &1423 &59&2858& 100\% \\
\midrule
6,000 & two-step  &  310 &  1116 & 36&1990&  95\%  \\
6,000 & allatonce  &  158 &968 &50&1998 & 96\% \\
6,000 & merged  &  158 &968 &53&1977& 96\% \\
\bottomrule
8,000 & two-step  &  283 &  889 & 36&1582&  89\%  \\
8,000 & allatonce  &  108 &714 &44&1581 & 91\% \\
8,000 & merged  &  108 &714 &44&1587& 90\% \\
\bottomrule
10,000 & two-step  &  251 &  757 & 40&1385&  82\%  \\
10,000 & allatonce  &  113 &615 &44&1383 & 83\% \\
10,000 & merged  &  113 &615 &43&1392& 82\% \\
\bottomrule
\end{tabular}
\end{table}
\begin{figure}
 \centering
 \includegraphics[width=1\linewidth]{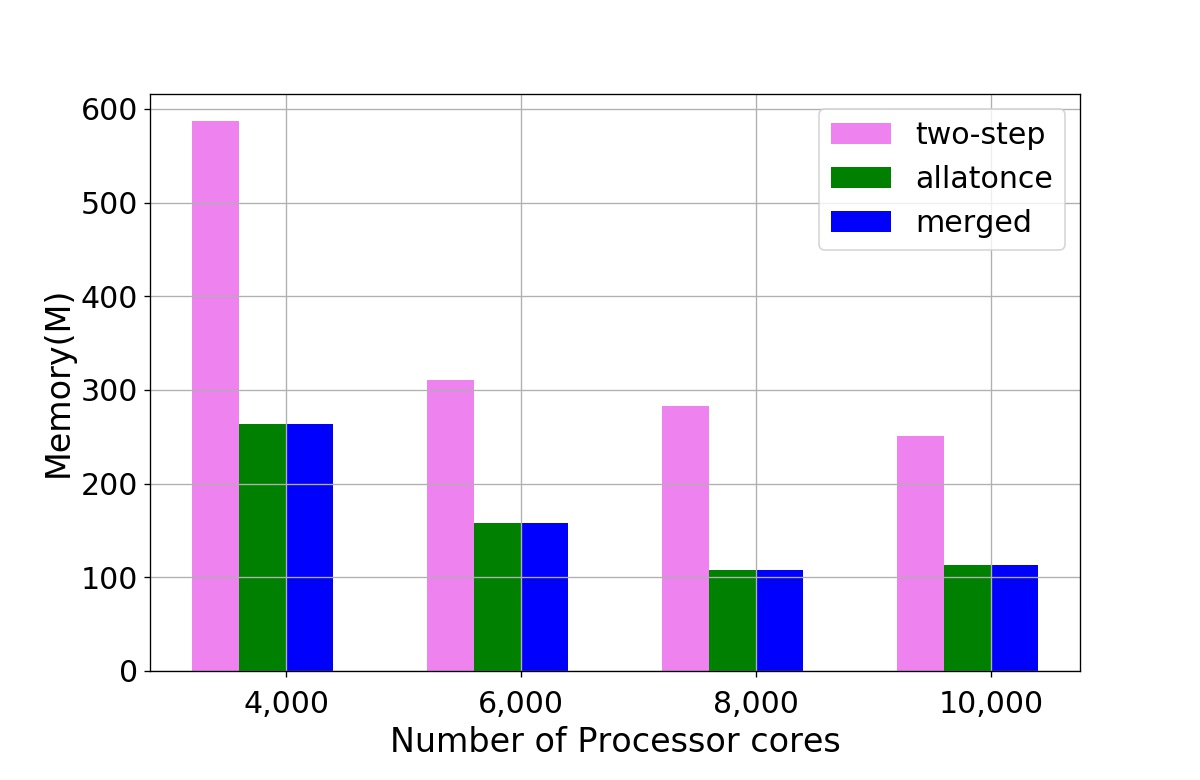} 
 \caption{Memory usages of different triple-product algorithms. \label{fig:s2memory}}
\end{figure}
The memory usage comparison for different triple product algorithms is also shown in Fig.~\ref{fig:s2memory}, where it is obvious that the memory used in the two-step method is twice as much as   used in the new algorithms for all processor counts. 

\begin{figure}
 \centering
 \includegraphics[width=1\linewidth]{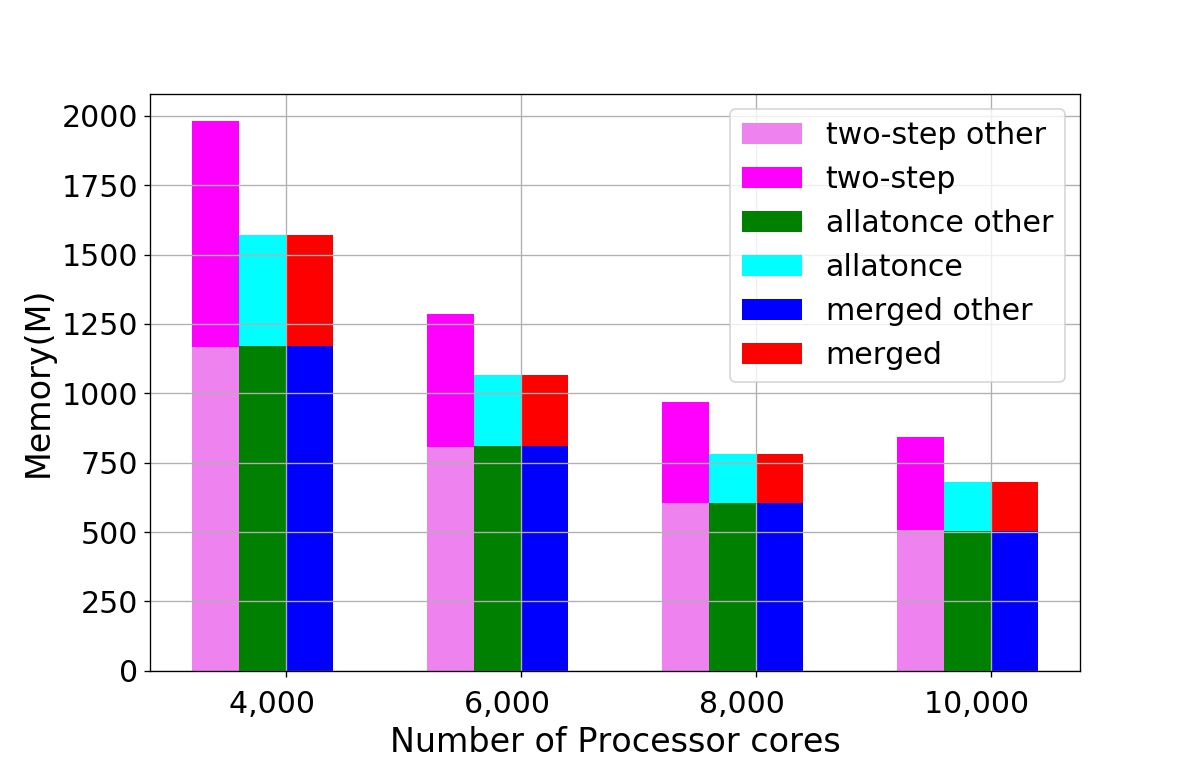} 
 \caption{Memory usages  on different triple-products and other part of the simulation. ``xxx other" represents the memory  used on the overall simulation except the triple products.  \label{fig:s2memorynofree}}
\end{figure}

\begin{table}
\scriptsize
\centering
\caption{Memory usages and compute times of different triple product algorithms with caching intermediate data on 4,000, 6,000, 8,000 and 10,000 processor cores. \label{tab:comparisonAllatonceNofree}}
\begin{tabular}{c c c c c c  c c c c}
\toprule
$np$  & algorithm&Mem & Mem$_T$ & Time & Time$_T$& EFF  \\
\midrule
4,000 & two-step  &  813 &  1981 & 48&2864&  100\%  \\
4,000 & allatonce  &  402 &1571 &61&2873 & 100\% \\
4,000 & merged  &  402 &1571 &59&2881& 100\% \\
\midrule
6,000 & two-step  &  478 &  1284 & 42&1969&  97\%  \\
6,000 & allatonce  &  256 &1066 &53&1997 & 96\% \\
6,000 & merged  &  256 &1066 &54&1999& 96\% \\
\bottomrule
8,000 & two-step  &  361 &  967 & 41&1586&  90\%  \\
8,000 & allatonce  &  176 &782 &45&1584 & 91\% \\
8,000 & merged  &  176 &782 &44&1611& 89\% \\
\bottomrule
10,000 & two-step  &  336 &  842 & 38&1373& 83\%  \\
10,000 & allatonce  &  180 &682 &47&1402 & 82\% \\
10,000 & merged  &  180 &682 &45&1393& 83\% \\
\bottomrule
\end{tabular}
\end{table}
In the previous test, the intermediate data  is free after the preconditioner setup. But in some applications, the preconditioner setup is repeated for each nonlinear iteration  and the corresponding triple products are also repeated.  In this case, we could choose to cache necessary  intermediate data, and do not need to do the symbolic calculations from the scratch for saving  compute time. In Table~\ref{tab:comparisonAllatonceNofree},  we concern on the amount of the memory required for all the algorithms  when we cache the intermediate  data. It is found that the memory is almost doubled, compared with that used without storing  temporary data. For example, at 4,000 processor cores, $813$ M memory is allocated for  the two-step method when caching the intermediate data, while it is $587$ M when we do not store temporary data.  For the new algorithms, it is also similar and the memory  usage is increased by $50\%$.  However, the memory usages of the new algorithms are still much lower than that used in the two-step method. More precisely, the memory usages of the new algorithms are half of that in the two-step method for all processor counts.  While the compute time spent on the triple products is a tiny portion of the total compute time, the corresponding memory usage takes a large chunk of the total memory.  This is shown in Fig.~\ref{fig:s2memorynofree}.  For example, at 10,000 processor cores, for the two-step method, the memory usage on the triple products takes $40\%$ of the total memory, and for the new algorithms, the triple-products account for $26\%$ of the total memory. It is exactly the motivation to optimize the memory usage in the triple products. 
\begin{figure}
 \centering
 \includegraphics[width=1\linewidth]{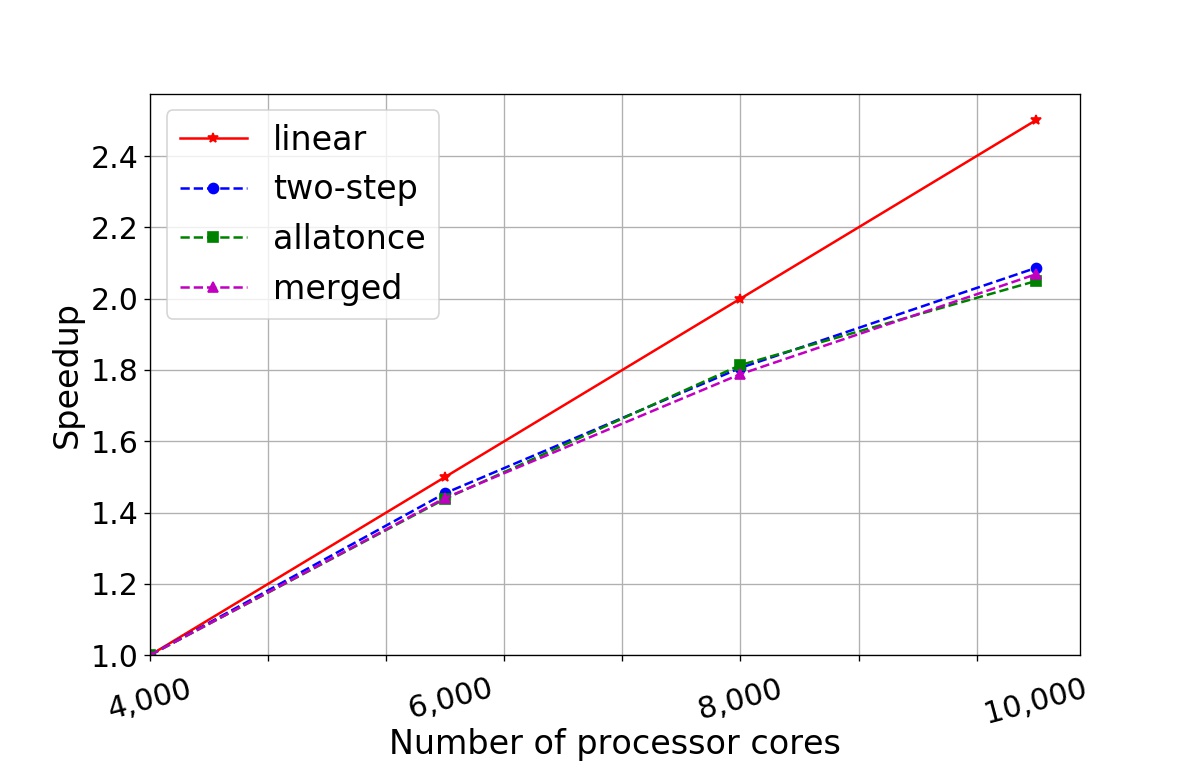} 
 \includegraphics[width=1\linewidth]{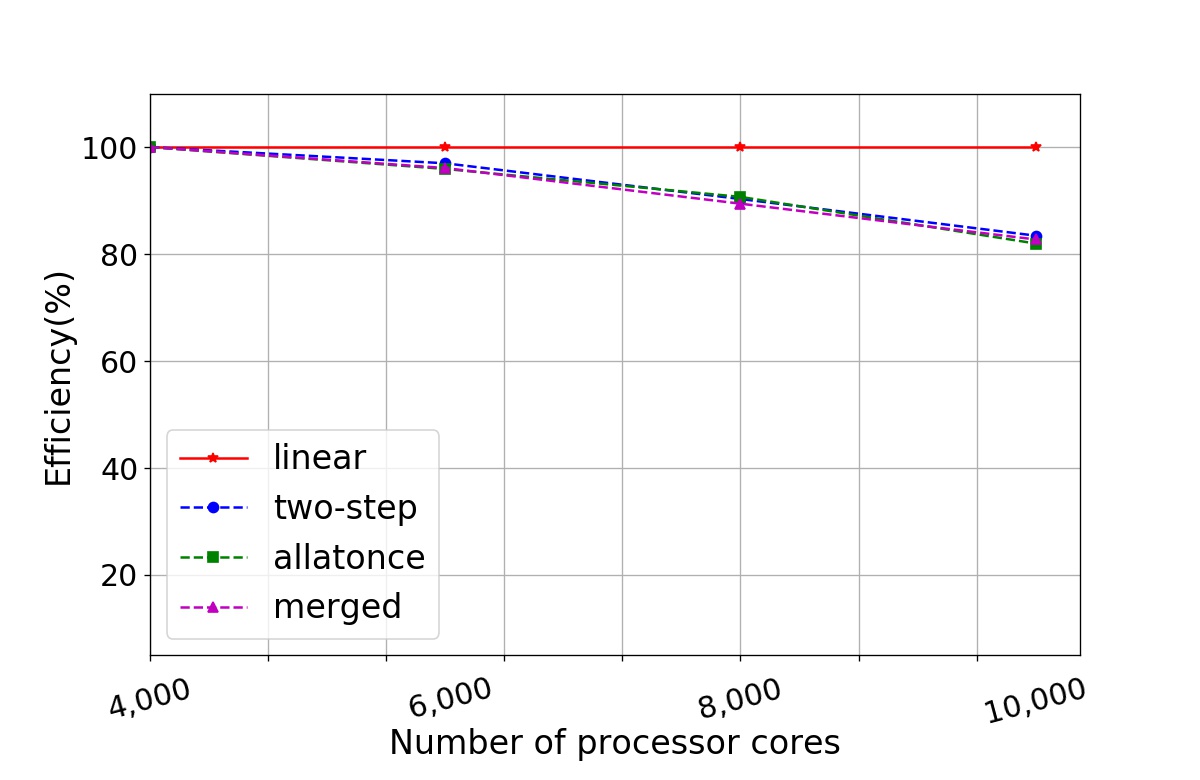} 
 \caption{Speedups and efficiencies of the entire simulation using different triple-product algorithms with cacheing  intermediate data. Top: speedups, bottom: parallel efficiencies.\label{fig:s2speedupsnofree}}
\end{figure}
Again, for all cases, the simulation is scalable in terms of the compute time and the memory usage.  The corresponding parallel efficiencies and speedups are drawn in Fig.~\ref{fig:s2speedupsnofree}.  We finally conclude that the new proposed algorithms including  the all-at-once and the merged all-at-once  use  similar  compute times as the two-step method does while they consume much less memory.

\section{Final remarks}
Two memory-efficient all-at-once algorithms are introduced, implemented and discussed for the sparse matrix triple products  in multigrid/multilevel methods.   The all-at-once triple product methods  based on hash tables form the output matrix   with one pass through the input matrices, where the first matrix-matrix multiplication  is carried out  using the row-wise approach and the second multiplication  is accomplished with an outer   product. For saving  memory and reducing communications, the all-at-once approach and its merged version  strategy do  not either  explicitly transpose the interpolation matrix  or  create any auxiliary matrices.    Compared with the traditional two-step method, the proposed new algorithms  use  much less memory while it is able to maintain the computational efficiency; e.g.,  it consumes only 10\% of  the memory  used in the two-step method for the model problem and 30\% for the realistic neutron 
transport simulation.  We have shown that the all-at-once algorithms are scalable in terms of the compute time and the memory usage for  the model problem with 27 billions of unknowns on supercomputer   with up to $32,768$ processor cores and  the realistic neutron transport problem with 2 billions of unknowns using $10,000$ processor cores.

\section*{Acknowledgments}
This manuscript has been authored by Battelle Energy Alliance, LLC under Contract No. DE-AC07-05ID14517 with the U.S. Department of Energy. The United States Government retains and the publisher, by accepting the article for publication, acknowledges that the United States Government retains a nonexclusive, paid-up, irrevocable, and worldwide license to publish or reproduce the published form of this manuscript, or allow others to do so, for United States Government purposes.

This research made use of the resources of the High-Performance Computing Center at Idaho National Laboratory, which is supported by the Office of Nuclear Energy of the U.S. Department of Energy and the Nuclear Science User Facilities under Contract No. DE-AC07-05ID14517.  This research also made use of the resources of the Argonne Leadership Computing Facility, which is a DOE Office of Science User Facility supported under Contract DE-AC02-06CH11357.

\bibliographystyle{sageh}
\bibliography{mpiptap_allatonce}

\end{document}